\documentclass[showpacs,preprintnumbers,amsmath,amssymb]{revtex4}

\usepackage{amsmath}
\usepackage{amssymb}
\usepackage{bm}
\usepackage{graphicx}
\usepackage{multirow}
\usepackage{textcomp}

\allowdisplaybreaks[1]

\begin{document}


\title{\boldmath QCD sum rules with spectral densities solved in inverse problems}
\author{Hsiang-nan Li}
\author{Hiroyuki Umeeda}
\affiliation{Institute of Physics, Academia Sinica,
Taipei, Taiwan 115, Republic of China}

\date{\today}

\begin{abstract}
We construct QCD sum rules for nonperturbative studies without assuming  
the quark-hadron duality for the spectral density at low energy on the hadron side.
Instead, both resonance and continuum contributions to the spectral density
are solved with the operator-product-expansion input on the quark side by treating sum rules as 
an inverse problem. This new formalism does not involve the continuum threshold, 
does not require the Borel transformation and stability analysis, 
and can be extended to extract properties of excited states. Taking the 
two-current correlator as an example, we demonstrate that the series of $\rho$ resonances 
can emerge in our formalism, and the decay constants 
$f_{\rho(770)} (f_{\rho(1450)},f_{\rho(1700)},f_{\rho(1900)})\approx$ 
0.22 (0.19, 0.14, 0.14) GeV for the masses $m_{\rho(770)} (m_{\rho(1450)},m_{\rho(1700)},m_{\rho(1900)})\approx$ 
0.78 (1.46, 1.70, 1.90) GeV are determined. We also show that the decay width $\Gamma_{\rho(770)}\approx 0.17$ 
GeV can be obtained by substituting a Breit-Wigner parametrization for the
$\rho(770)$ pole on the hadron side. 
It is observed that quark condensates of dimension-six on the quark side are crucial
for establishing those $\rho$ resonances. Handling the conventional sum rules with the 
duality assumption as an inverse problem, we find that the multiple pole sum rules widely 
adopted in the literature do not describe the $\rho$ excitations reasonably.
The precision of our theoretical outcomes can be 
improved systematically by including higher-order and higher-power corrections on the quark side. 
Broad applications of this formalism to abundant low energy QCD observables are expected.

\end{abstract}


%
%
%

\maketitle

%
%
%

\section{INTRODUCTION}

QCD sum rules have become one of the major nonperturbative approaches to 
low energy hadronic processes, since they were proposed decades ago \cite{SVZ}.
This approach relies heavily on the assumption of the quark-hadron duality for the
spectral density on the hadron side in a low energy region, whose theoretical uncertainty 
is difficult to control and quantify. The Borel transformation is applied to suppress  
model dependent continuum contributions on the hadron side and 
higher power corrections on the quark side. Nevertheless, 
the typical scale of a Borel mass may not be large enough for justifying 
the desired suppression. The choice of the continuum threshold is a bit arbitrary,
though the stability criterion, ie., the existence of the so-called "sum rule window" under 
the variation of the continuum threshold and Borel mass has been imposed.
However, the above prescriptions are quite discretionary \cite{Coriano:1993yx},
such that strong dependence on 
the continuum threshold and Borel mass is not avoidable.
Besides, one usually invokes the vacuum saturation hypothesis (factorization) to 
replace higher dimension condensates by products of lower dimension ones on the quark side, 
which also causes uncertainty. Therefore, there has been concern on the rigorousness and 
predictive power of QCD sum rules \cite{Leinweber:1995fn,Gubler:2010cf}.

In this paper we will handle QCD sum rules in a different way, attempting to resolve the
aforementioned difficulty to some extent.
The spectral density on the hadron side of a sum rule, 
including both resonance and continuum contributions, is regarded as an unknown. The 
operator product expansion (OPE) on the quark side is calculated in the standard
way. A sum rule is then treated as an inverse problem, in which the unknown 
(source distribution) is solved from the OPE input 
(potential observed outside the distribution). This formalism does not involve
the continuum threshold, because the continuum can be a smooth distribution not related to
the perturbative spectral density. It does not require a Borel transformation to
suppress the continuum contribution, which will be solved from the inverse problem.
The suppression on the higher power corrections can be achieved by considering
the input in the deep euclidean region. Once the unknown spectral density is
solved directly, the stability criterion for a conventional sum 
rule is not necessary. Certainly, the Borel transformation can be applied to our
formalism, but it will be verified that results from the versions with and without this
transformation are similar.

It has been known that an inverse problem is ill-posed and allows the existence of multiple 
solutions. We will show that the existence of multiple solutions grants the extension of our
formalism to studies of excited states, which impose a challenge to conventional
sum rules. Taking the two-current correlator as an example, we demonstrate how to obtain
the masses and decay constants of the $\rho$ resonances. We first fix the correction
to the vacuum saturation hypothesis for higher dimension condensates on the quark side from the input
of the ground state $\rho(770)$ mass, and determine the $\rho(770)$ meson decay constant  
by solving the sum rule as an inverse problem. The lower state observables are then adopted
as inputs to extract properties of higher states one by one. A series of (radial) 
excitations can be probed systematically following the above strategy.
The masses $m_{\rho(770)} (m_{\rho(1450)},m_{\rho(1700)},m_{\rho(1900)})\approx$ 0.78 (1.46, 1.70, 1.90) 
GeV and the decay constants $f_{\rho(770)} (f_{\rho(1450)},f_{\rho(1700)},f_{\rho(1900)})\approx$ 
0.22 (0.19, 0.14, 0.14) GeV are extracted. Besides, the decay width $\Gamma_{\rho(770)}\approx 0.17$ 
GeV is also derived by substituting a Breit-Wigner parametrization for the $\rho(770)$ pole on the 
hadron side. To understand how the nonperturbative condensates influence the appearance of the $\rho$
resonance, we examine the impacts from power corrections of various dimensions.
It is found that the quark condensate of dimension-six plays a crucial role for 
establishing the $\rho(770)$ state. 

Properties of excited states have been investigated in conventional QCD sum rules
by employing the double pole plus continuum model for a spectral density 
\cite{SVVZ,NOS,Krasnikov:1981vw,Krasnikov:1982ea,Bakulev:1998pf,Pimikov:2013usa}.
The second pole for the excited state was put in by hand, and ad hoc prescriptions for
choosing an appropriate continuum threshold have to be postulated \cite{MaiordeSousa:2012vv},
such as the lower bound of the continuum threshold being set to the excited state mass plus 100 MeV.
Treating the above conventional sum rules as an inverse problem, we explicitly show that there are no
signs for excited $\rho$ resonances, once the quark-hadron duality is assumed. 
This investigation casts doubt on the multiple pole QCD sum rules, which have been widely adopted 
in the literature. Our formalism is close to the Bayesian approach to QCD sum rules \cite{Gubler:2010cf},
in which the specific form of the spectral density was not assumed, but 
derived using the maximum entropy method. 
Though it is possible to explore the existence of excited states by applying 
this method to sum rules, at least its application to the nucleon 
mass spectrum has not been successful \cite{Ohtani:2012ps}. We suspect that
the failure is attributed to the ill-posed essence of an inverse problem, which
makes difficult searching for correct excitations from many allowed 
solutions without any specific parametrization for the spectral density. Hence, our work 
provides a justified and practical approach to studies of excited states based on QCD sum rules.

The rest of the paper is organized as follows.
In Sec.~II we construct our formalism starting from the dispersion relation
for a two-current correlator. The distinction from conventional QCD sum rules, 
namely, no assumption of the quark-hadron duality, is highlighted. We elaborate the 
extractions of the masses and decay constants of the series of $\rho$ resonances by 
solving the sum rules as an inverse problem in Sec.~III, starting with the $\rho(770)$ 
meson mass, which is used to fix the factorization violation parameter associated with 
the dimension-six condensate. We end the search for the $\rho$ excitations at $\rho(2000)$
in our formalism, for which theoretical and experimental studies are still rare.
The conventional multiple pole sum rules with the duality assumption
are also solved in a similar way to confirm our concern about their applications
to excited states. Section IV contains the conclusion and outlooks.

\section{FORMALISM}

\begin{figure}
\includegraphics[scale=0.35]{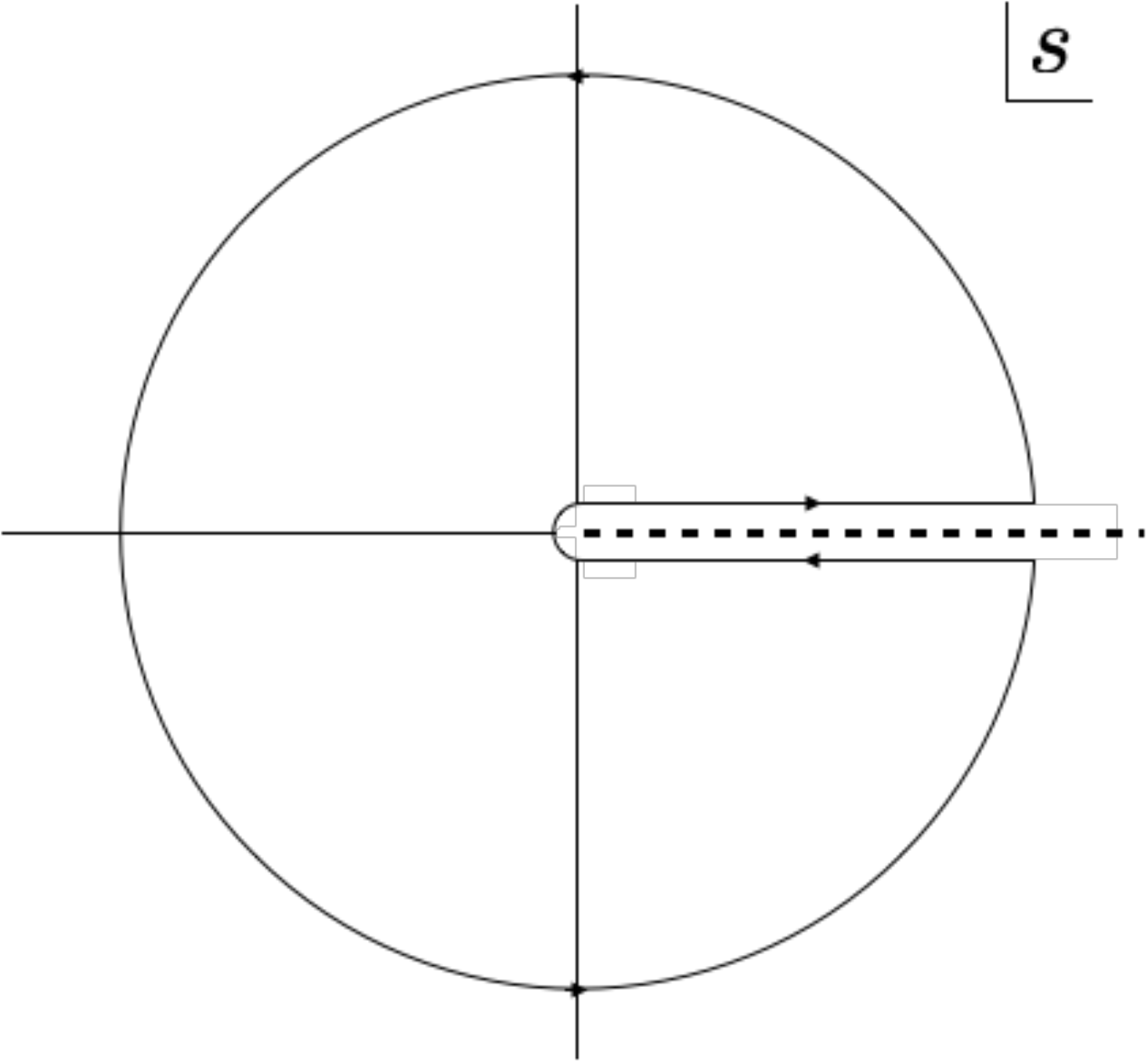}
\caption{\label{fig0}
Contour on the complex $s$ plane.}
\end{figure}

The series of $\rho$ resonances is one of the first objects analyzed in QCD sum rules, through which  
we formulate our approach and demonstrate its application.
We first briefly recollect the idea of conventional QCD sum rules, starting with the 
two-point correlator
\begin{eqnarray}
\Pi_{\mu\nu}(q^2)=i\int d^4xe^{iq\cdot x}
\langle 0|T[J_\mu(x)J_\nu(0)]|0\rangle=(q_\mu q_\nu-g_{\mu\nu}q^2)\Pi(q^2),\label{cur}
\end{eqnarray}
for the current $J_\mu=(\bar u\gamma_\mu u-\bar d\gamma_\mu d)/\sqrt{2}$.
The vacuum polarization function $\Pi(q^2)$ obeys the identity 
\begin{eqnarray}
\Pi(q^2)=\frac{1}{2\pi i}\oint ds\frac{\Pi(s)}{s-q^2},\label{di1}
\end{eqnarray}
where the contour, depicted in Fig.~\ref{fig0}, consists of two pieces of horizontal lines above 
and below the positive horizontal axis, ie., the branch cut, and a circle of large radius $R$.
For $s$ far away from physical poles, the perturbative evaluation of $\Pi(s)$ is reliable,
so the right hand side of Eq.~(\ref{di1}) can be written as
\begin{eqnarray}
\frac{1}{2\pi i}\oint ds\frac{\Pi(s)}{s-q^2}=
\frac{1}{\pi}\int_{s_i}^\Lambda ds\frac{{\rm Im}\Pi(s)}{s-q^2}
+\frac{1}{\pi}\int_\Lambda^R ds\frac{{\rm Im}\Pi^{\rm pert}(s)}{s-q^2}
+\frac{1}{2\pi i}\int_C ds\frac{\Pi^{\rm pert}(s)}{s-q^2},\label{di2}
\end{eqnarray}
where $s_i$ in the first integral denotes the threshold for the nonvanishing spectral density
${\rm Im}\Pi(s)$, the numerator in the second integrand has been replaced by the perturbative spectral 
density ${\rm Im}\Pi^{\rm pert}(s)$ for a sufficiently large separation scale $\Lambda$,
and $C$ in the third integral represents the large circle of radius $R$. The spectral density
${\rm Im}\Pi(s)$ in the first integrand, involving nonperturbative dynamics from the low $s$ region, 
will be determined later. The perturbative function $\Pi^{\rm pert}(s)$ in the third integral
receives only the perturbative QCD contribution.

For $q^2$ in the deep Euclidean region, the 
OPE of $\Pi(q^2)$ is reliabe, and we have $\Pi^{\rm pert}(q^2)$ \cite{SVZ}
for the left hand side of Eq.~(\ref{di1})
\begin{eqnarray}
\Pi^{\rm pert}(q^2)=\frac{1}{2\pi i}\oint ds\frac{\Pi^{\rm pert}(s)}{s-q^2}+
\frac{1}{12\pi}\frac{\langle\alpha_sG^2\rangle}{(q^2)^2}+
2\frac{\langle m_q \bar q q\rangle}{(q^2)^2} +\frac{224\pi}{81}
\frac{\kappa \alpha_s\langle \bar q q\rangle^2}{(q^2)^3},\label{di3}
\end{eqnarray}
up to the dimension-six condensate, ie., up to the power correction of $1/(q^2)^3$. 
In the above expression $\langle G^2\rangle$ is the gluon condensate, 
$m_q$ is a quark mass, and the parameter $\kappa=2$-4 \cite{CDK,SN95,SN09} is introduced to 
quantify the violation in the factorization of the four-quark condensate $\langle (\bar q q)^2\rangle$ 
into the product of $\langle \bar q q\rangle$. The first term on the right hand side, 
collecting higher order corrections, has been expressed
as the integral of the perturbative function $\Pi^{\rm pert}(s)$ along the contour in Fig.~\ref{fig0}. 

The equality of Eq.~(\ref{di2}) on the hadron side and Eq.~(\ref{di3}) on the quark side
leads to 
\begin{eqnarray}
\frac{1}{\pi}\int_{s_i}^\Lambda ds\frac{{\rm Im}\Pi(s)}{s-q^2}=
\frac{1}{\pi}\int_{s_i}^\Lambda ds\frac{{\rm Im}\Pi^{\rm pert}(s)}{s-q^2}+
\frac{1}{12\pi}\frac{\langle\alpha_sG^2\rangle}{(q^2)^2}+
2\frac{\langle m_q \bar q q\rangle}{(q^2)^2} +\frac{224\pi}{81}
\frac{\kappa \alpha_s\langle \bar q q\rangle^2}{(q^2)^3},\label{di4}
\end{eqnarray}
where the contributions of the perturbative function $\Pi^{\rm pert}(s)$ in the regions 
away from physical poles have cancelled from both sides, and only the perturbative spectral density
\begin{eqnarray}
{\rm Im}\Pi^{\rm pert}(q^2)=\frac{1}{4\pi}\left(1+\frac{\alpha_s}{\pi}\right)\equiv a\pi,
\end{eqnarray}
along the branch cut remains. Equation~(\ref{di4}) is a result of the dispersion relation 
for the function $\Pi(q^2)$.

The next step is to parametrize the nonperturbative spectral density ${\rm Im}\Pi(s)$
on the left hand side of Eq.~(\ref{di4}). The translational
invariance and the integration over the coordinate $x$ in Eq.~(\ref{cur}) gives
\begin{eqnarray}
2{\rm Im}\Pi_{\mu\nu}(q^2)=\sum_n\langle 0|J_\mu|n\rangle\langle n|J_\nu|0\rangle
d\Phi_n(2\pi)^4\delta(q-p_n),\label{i}
\end{eqnarray}
for $q^2>0$, in which $d\Phi_n$ and $p_n$ represent the phase space and the momentum of
the intermediate state $|n\rangle$, respectively. The ground state for $|n\rangle$ is a neutral 
vector of the mass $m_V$ and the polarization vector $\epsilon$, 
which defines the decay constant $f_V$ via the matrix element
\begin{eqnarray}
\langle 0|J_\mu|V^\lambda\rangle=f_Vm_V\epsilon^\lambda_\mu.\label{v}
\end{eqnarray}
The substitution of Eq.~(\ref{v}) into Eq.~(\ref{i}) yields
\begin{eqnarray}
{\rm Im}\Pi(q^2)=\pi f_V^2\delta(q^2-m_V^2)+\pi\rho^h(q^2)\theta(q^2-s_h),\label{imp}
\end{eqnarray}
where the first term is a consequence of the narrow width approximation,
and the second term describes the contribution from higher excitations with $s_h$ being their 
threshold. It has been assumed that the widths of excited states become broader, so 
their contributions can be parametrized as a continuous spectral density function 
$\rho^h(q^2)$.

The key of QCD sum rules is the quark-hadron duality, which assumes
that the spectral density $\rho^h(s)$ is related to the perturbative density 
${\rm Im}\Pi^{\rm pert}(s)$ as $s$ is higher than some scale $s_0>s_h$ by
\begin{eqnarray}
\rho^h(s)=\frac{1}{\pi}{\rm Im}\Pi^{\rm pert}(s)\theta(s-s_0),
\label{non}
\end{eqnarray}
referred to the local duality, or by
\begin{eqnarray}
\int_{s_h}^\Lambda ds\frac{\rho^h(s)}{s-q^2}=
\frac{1}{\pi}\int_{s_0}^\Lambda ds\frac{{\rm Im}\Pi^{\rm pert}(s)}{s-q^2},\label{glo}
\end{eqnarray}
referred to the global duality. The threshold $s_0$ is around
1 GeV$^2$, so the duality can hardly hold at such a low scale 
\cite{Boito:2012nt,Rodriguez-Sanchez:2016jvw}. Obviously, the quark-hadron 
duality is a major source of theoretical uncertainty, which is not easy to control.

The Borel transformation 
\begin{eqnarray}
{\hat B}_M\equiv \lim_{\begin{array}{c}Q^2,n\to\infty\\Q^2/n=M^2\end{array}}
\frac{1}{(n-1)!}(Q^2)^n\left(-\frac{d}{dQ^2}\right)^n,
\end{eqnarray}
with $Q^2\equiv -q^2$, is then employed to suppress the continuum contribution on the 
hadron side, which has been related to the perturbative spectral density via the duality assumption,
and to improve the OPE on the quark side. 
Inserting Eqs.~(\ref{imp}) and (\ref{non}) (or (\ref{glo})) into Eq.~(\ref{di4}), we derive the 
conventional sum rule under the Borel transformation \cite{SVZ}
\begin{eqnarray}
f_V^2e^{-m_V^2/M^2}=\frac{1}{\pi}\int_{s_i}^{s_0} ds{\rm Im}\Pi^{\rm pert}(s)e^{-s/M^2}
+\frac{1}{12\pi}\frac{\langle\alpha_sG^2\rangle}{M^2}
+2\frac{\langle m_q \bar q q\rangle}{M^2} -\frac{112\pi}{81}
\frac{\kappa \alpha_s\langle \bar q q\rangle^2}{M^4}.\label{sumb}
\end{eqnarray}
It is seen that the suppression on the higher power corrections with the typical 
$M\sim O(1)$ GeV and by the additional factors $1/(k-1)!$ for the $1/(q^2)^k$ term, 
$k\le 3$, is not effective actually. The prescription for a sum rule calculation
is to tune the threshold $s_0$ in the above formula, such that the value of $f_V$ is stable 
against the variation of the Borel mass $M$ in a maximal window of $M$.
This prescription introduces theoretical uncertainly, 
especially when the sum rule window does not exist \cite{Coriano:1993yx}.

An alternative interpretation for the equality of the hadron and quark sides
in Eq.~(\ref{di4}) is that there exist multiple
solutions to Eq.~(\ref{di1}): the right hand side of Eq.~(\ref{di2}), 
which contains the nonperturbative spectral density ${\rm Im}\Pi(s)$ in the first term, can be regarded as
a nonperturbative solution to Eq.~(\ref{di1}), while the right hand side of Eq.~(\ref{di3})
can be regarded as a perturbative solution. Motivated by the above viewpoint, 
we propose to handle QCD sum rules as an inverse problem, for which multiple solutions
exist naturally. First, the spectral density is written as the superposition of the pole 
and continuum contributions 
\begin{eqnarray}
{\rm Im}\Pi(q^2)=\pi f_V^2\delta(q^2-m_V^2)+\pi\rho^h(q^2),\label{imp1}
\end{eqnarray}
where the threshold $s_h$ in Eq.~(\ref{imp}), introduced in conventional sum rules to 
characterize the continuum region, does not appear. As mentioned before, excited 
states tend to have broader widths, and the 
transition from the resonance to continuum region should be smooth. Hence, the 
second term in Eq.~(\ref{imp1}) behaves more like a ramp function \cite{Kwon:2008vq} in general,
taking a value as $q^2 > s_i$, instead of like a step function. The unknown function
$\pi\rho^h(q^2)$ can be approximated by the perturbative spectral density
${\rm Im}\Pi^{\rm pert}(q^2)$ reliably as $q^2$ is great than some large 
separation scale $\Lambda$ as shown in Eq.~(\ref{di2}). Strictly speaking, this approximation 
is also based on the local quark-hadron duality, but it is not the concerned duality assumption 
in conventional QCD sum rules around the threshold $s_0\approx 1$ GeV$^2$, and the duality 
violation above the large $\Lambda$ is expected to be minor.

Inserting Eq.~(\ref{imp1}) into the left hand side of Eq.~(\ref{di4}), we write
\begin{eqnarray}
& &\frac{f_V^2}{m_V^2-q^2}+\int_{0}^\Lambda ds\frac{\rho^h(s)}{s-q^2}
=\omega(q^2),\nonumber\\
& &\omega(q^2)= a\ln\frac{q^2-\Lambda}{q^2}+
\frac{1}{12\pi}\frac{\langle\alpha_sG^2\rangle}{(q^2)^2}+
2\frac{\langle m_q \bar q q\rangle}{(q^2)^2} +\frac{224\pi}{81}
\frac{\kappa \alpha_s\langle \bar q q\rangle^2}{(q^2)^3}
\label{sum1}
\end{eqnarray}
in which the threshold $s_i=4m_\pi^2$ with the pion mass $m_\pi$ 
has been set to zero, and the OPE input $\omega(q^2)$, equal to 
the right hand side of Eq.~(\ref{di4}),
is calculable as a standard OPE. The sum rule is then turned into an inverse
problem, where the unknowns $m_V$, $f_V$ and $\rho^h(s)$ are solved with the 
OPE input $\omega(q^2)$. 
The suppression on the uncertain continuum contribution, 
which will be solved directly, is not necessary. The suppression on the higher power corrections 
can be easily achieved by considering the input $\omega(q^2)$ at large $|q^2|$.
Applying the Borel transformation to Eq.~(\ref{sum1}), we get
\begin{eqnarray}
& &\frac{f_V^2}{M^2}e^{-m_V^2/M^2}+\frac{1}{M^2}\int_{0}^\Lambda ds\rho^h(s)e^{-s/M^2}={\hat\omega}(M^2),\nonumber\\
& &{\hat\omega}(M^2)\equiv {\hat B}_M\omega(q^2)=a(1-e^{-\Lambda/M^2})+\frac{1}{12\pi}\frac{\langle\alpha_sG^2\rangle}{(M^2)^2}+
2\frac{\langle m_q \bar q q\rangle}{(M^2)^2} -\frac{112\pi}{81}
\frac{\kappa \alpha_s\langle \bar q q\rangle^2}{(M^2)^3}.\label{ivb}
\end{eqnarray}
As demonstrated in the next section, both versions, Eqs.~(\ref{sum1}) and (\ref{ivb}), give similar solutions to 
the unknowns. Therefore, we postulate that the Borel transformation is not crucial for the present  
formalism.

Note that an inverse problem is usually ill-posed, and the ordinary discretization method to solve 
a Fredholm integral equation does not work. The best fit method proposed in \cite{Li:2020xrz}
may be the most transparent way to reveal the existence of multiple solutions in this case.
To facilitate the numerical analysis, we expand the spectral density function 
$\rho^h(y)\equiv\rho^h(s=y\Lambda)$ in Eq.~(\ref{sum1}) in a
series of Legendre polynomials
\begin{eqnarray}
\rho^h(y)=b_0P_0(2y-1)+b_1P_1(2y-1)+b_2P_2(2y-1)+b_3P_3(2y-1)+\cdots,\label{exp}
\end{eqnarray}
with 
\begin{eqnarray}
P_0(y)=1,\;\;\;P_1(y)=y,\;\;\;P_2(y)=\frac{1}{2}(3y^2-1),\;\;\;
P_3(y)=\frac{1}{2}(5y^3-3y).
\end{eqnarray}
Other bases of orthogonal functions, such as the trigonometric functions, can serve the purpose 
equally well. The boundary conditions $\rho^h(0)=0$ and $\rho^h(1)=a$ (equal to the perturbative density
at $s=\Lambda$) impose the constraints
\begin{eqnarray}
b_2=\frac{a}{2}-b_0,\;\;\;b_3=\frac{a}{2}-b_1.
\end{eqnarray}
It will be verified that the expansion up to $P_3(y)$, with the converging coefficients $b_i$, 
is sufficient. We will solve Eq.~(\ref{sum1}) by tuning 
$\Lambda$, $m_V$, $f_V$, $b_0$ and $b_1$ to minimize the difference between its two sides.

\section{APPLICATIONS}

In this section we extract the observables associated with the series of $\rho$ resonances from 
our formalism. It is notoriously difficult to solve a Fredholm integral equation like Eq.~(\ref{sum1}).
We have found that the best fit method may be the most transparent way to probe
multiple solutions of a Fredholm equation, which has been applied to the 
explanation of the $D$ meson mixing parameters \cite{Li:2020xrz} and the determination of the 
hadronic vacuum polarization contribution to the muon anomalous magnetic moment \cite{LU}. 
The following OPE parameters \cite{Wang:2016sdt,Narison:2014wqa} and the strong 
coupling evaluated at the scale of 1 GeV are adopted:
\begin{eqnarray}
& &
\Lambda_{\rm QCD} = 0.353\;{\rm GeV},\;\;
\langle m_q\bar q q\rangle = 0.007\times(-0.246)^3\;{\rm GeV}^4,\;\;
\langle\alpha_sGG\rangle=0.07\; {\rm GeV}^4,\nonumber\\
& &\alpha_s\langle \bar q q\rangle^2 = 1.49\times 10^{-4}\;{\rm GeV}^6,\;\;\alpha_s=0.5.
\end{eqnarray}
We consider the input $\omega(q^2)$ from an appropriate range of $q^2$ in the Euclidean
region, in which 20 points $q^2_i$ are selected,
and then search for the set of parameters $\Lambda$, $m_V$, $f_V$, $b_0$ and $b_1$ 
that minimizes the residual sum of square (RSS) 
\begin{eqnarray}
\sum_{i=1}^{20} \left|\frac{f_V^2}{m_V^2-q_i^2}
+\int_{0}^\Lambda ds\frac{\rho^h(s)}{s-q_i^2}-\omega(q^2_i)\right|^2.\label{dev}
\end{eqnarray}
Such a set of parameters approximates a solution of the Fredholm equation~(\ref{sum1}).
A similar RSS can be defined for the sum rule in Eq.~(\ref{ivb}) under the Borel transformation,
for which the input ${\hat\omega}(M^2)$ is selected from an appropriate range of $M^2>0$.
We have tested the number of input points from 20 to 500, and confirmed that solutions do not alter. 

\subsection{Ground State}

\begin{figure}
\includegraphics[scale=0.35]{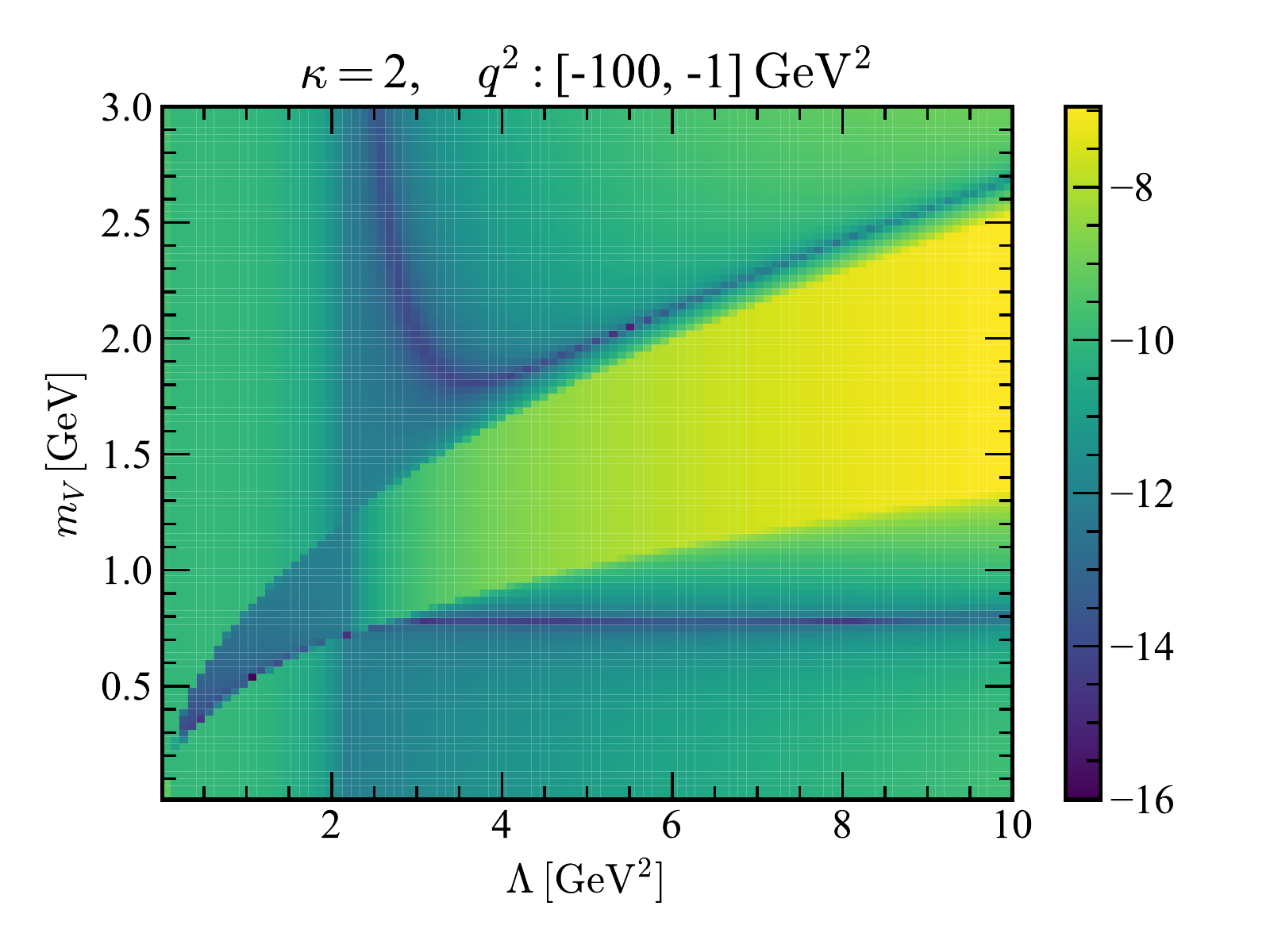}
\includegraphics[scale=0.35]{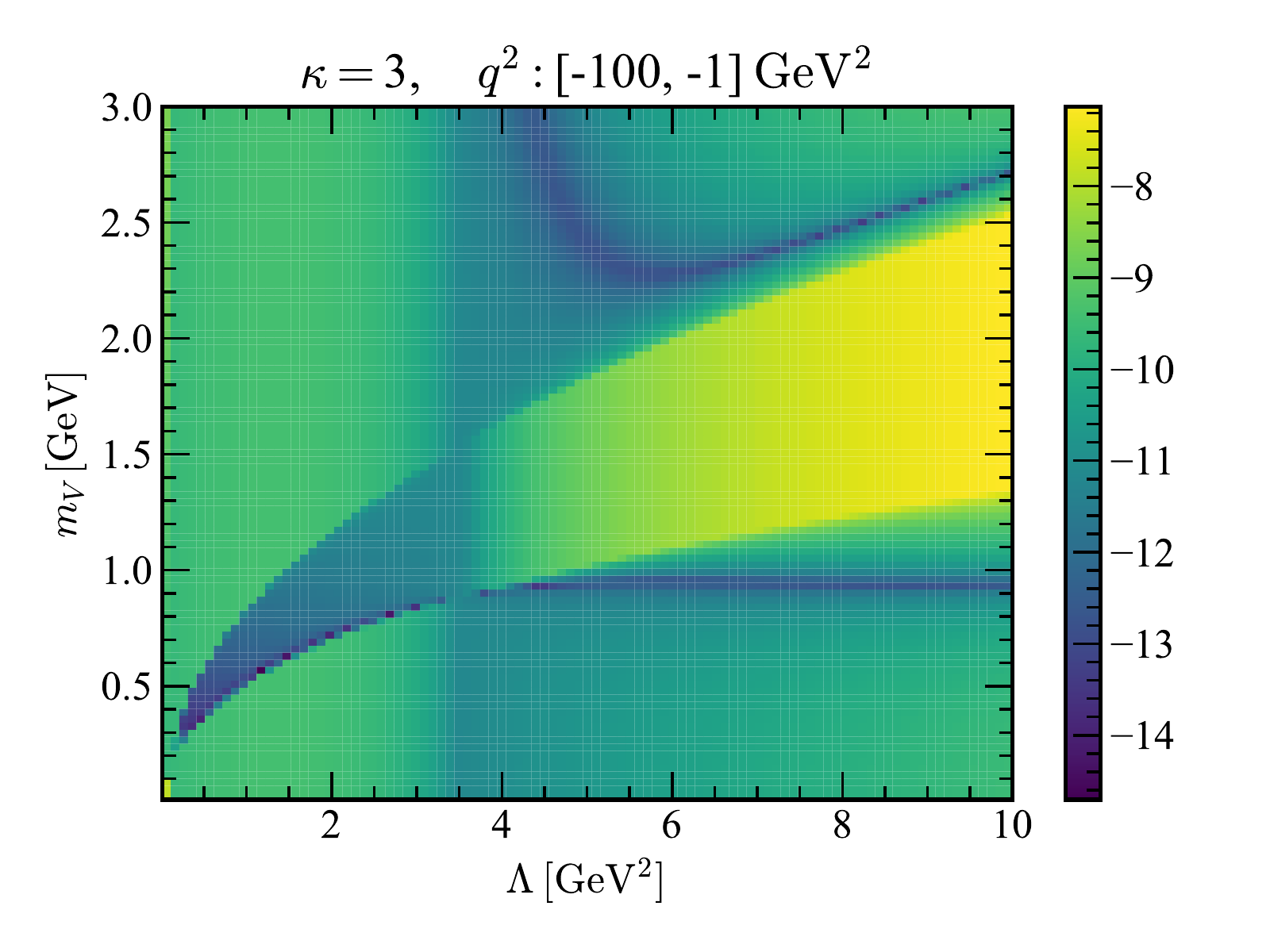}
\includegraphics[scale=0.35]{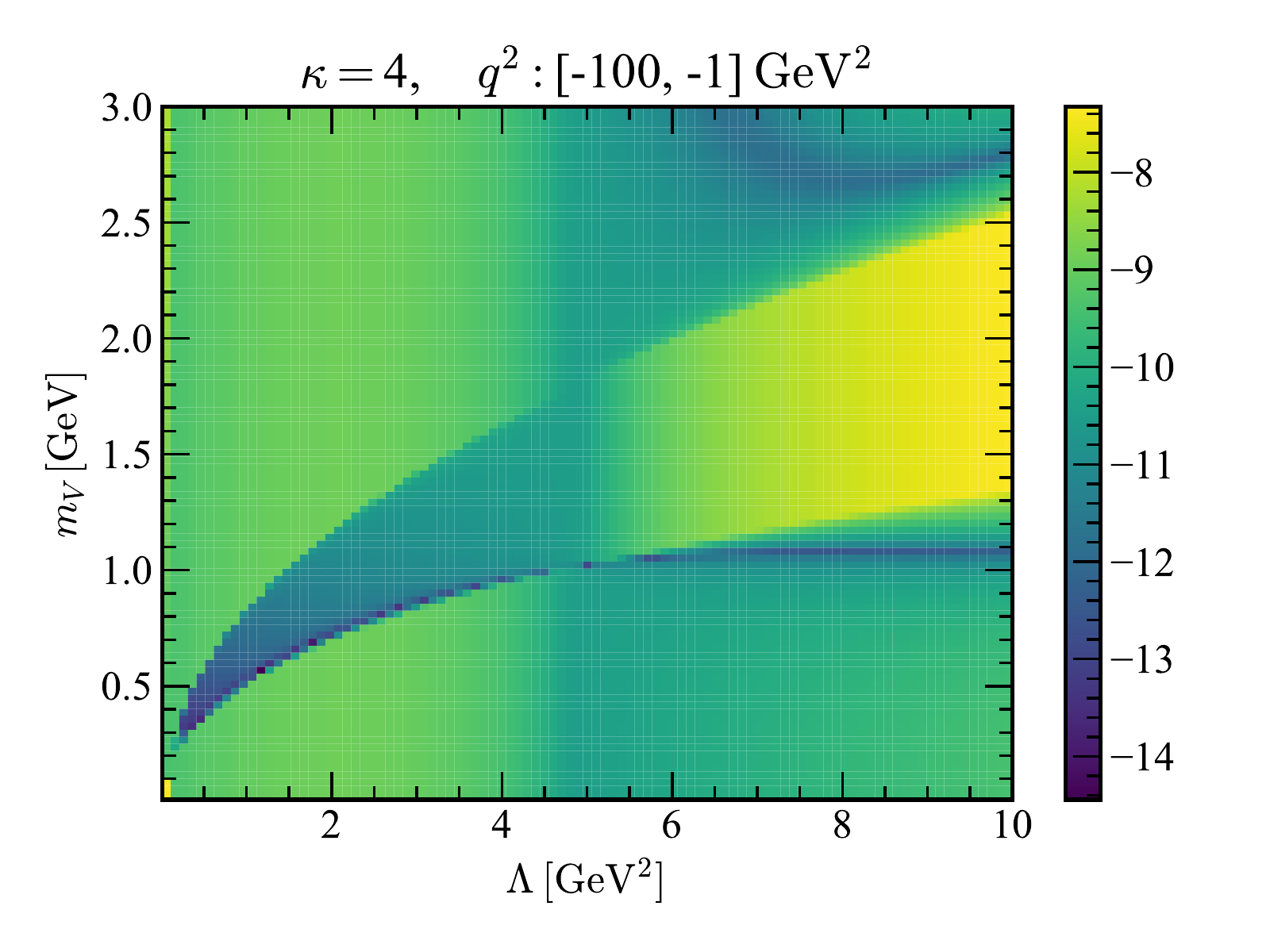}
\includegraphics[scale=0.35]{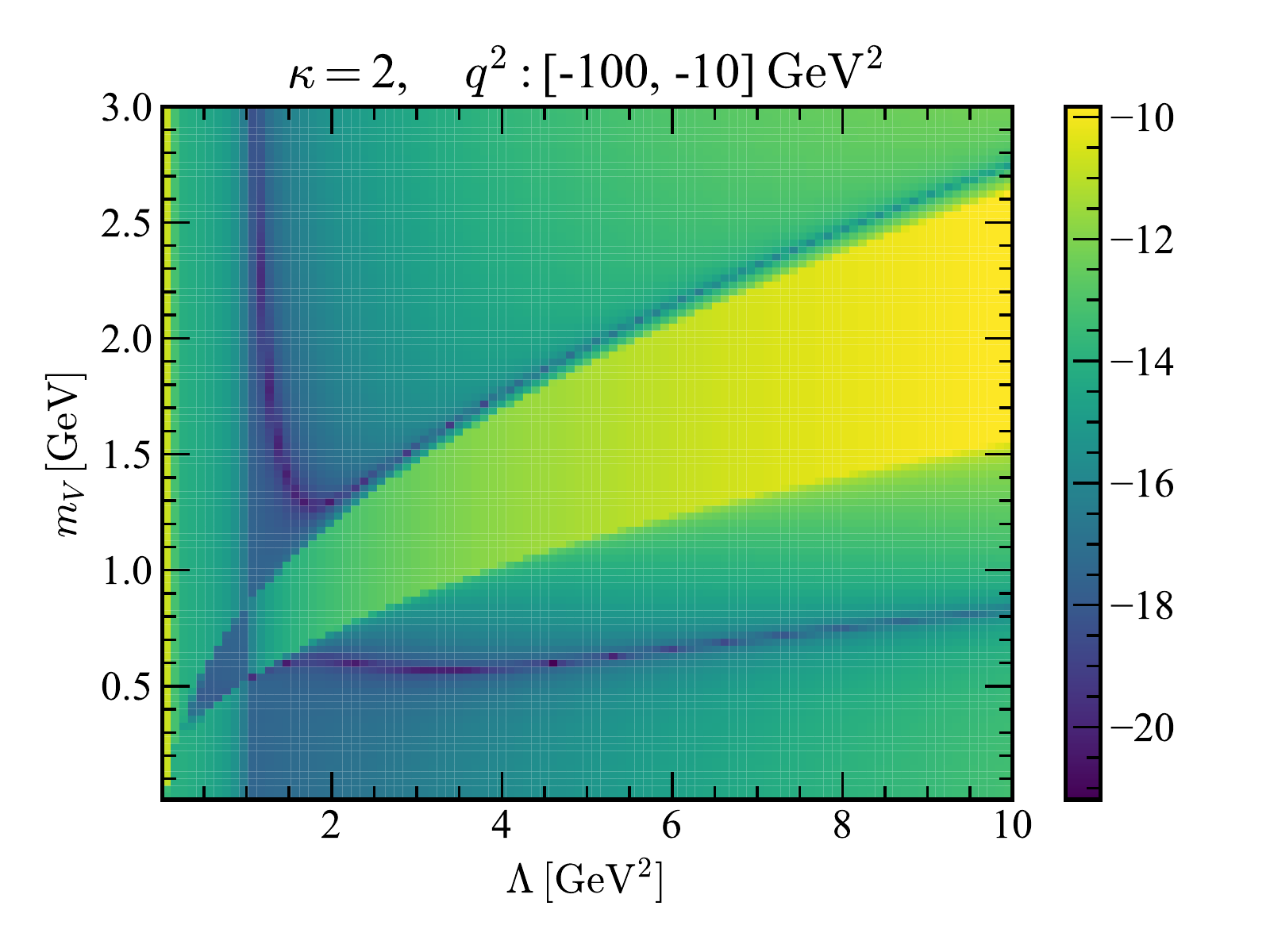}
\includegraphics[scale=0.35]{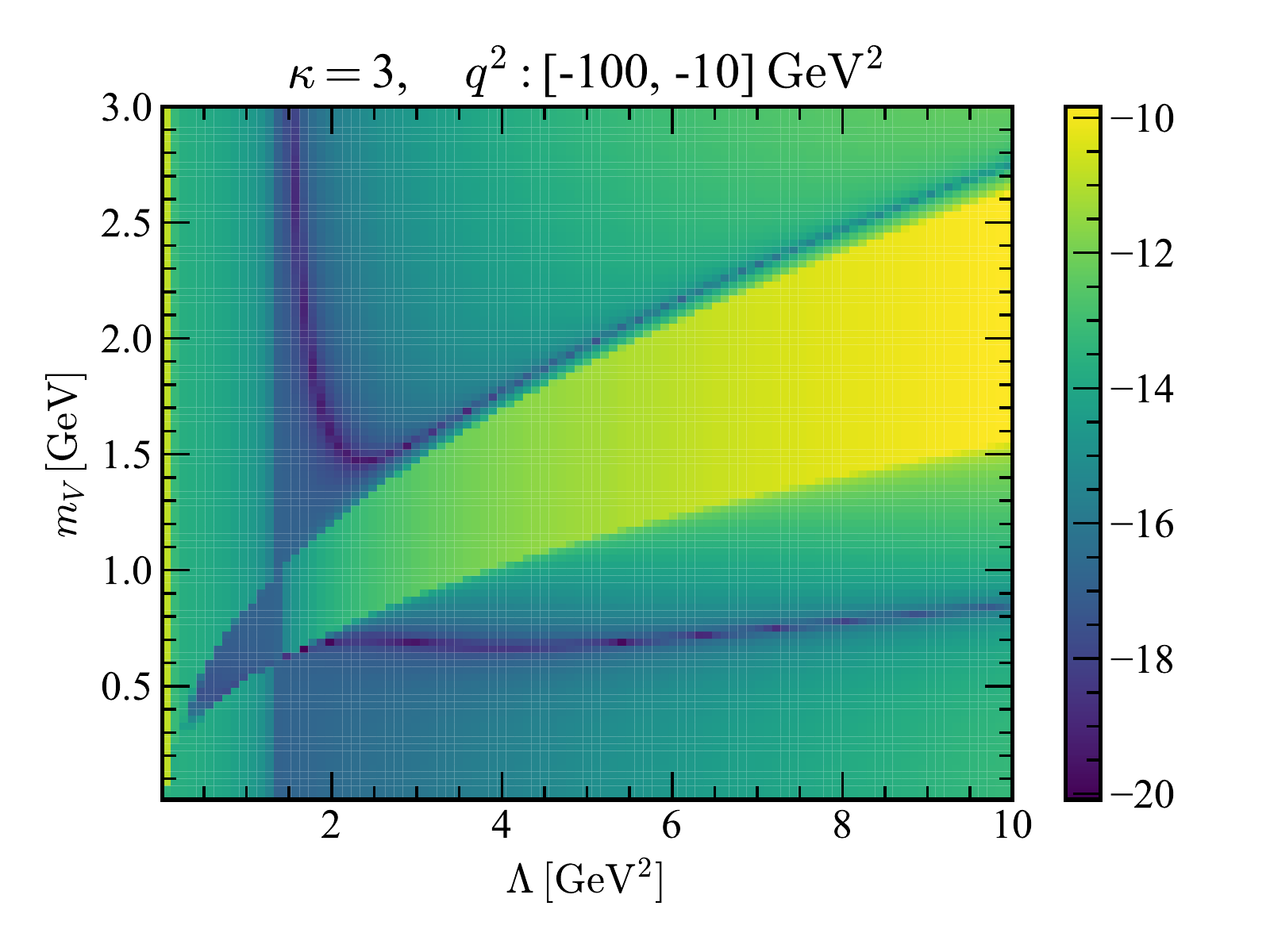}
\includegraphics[scale=0.35]{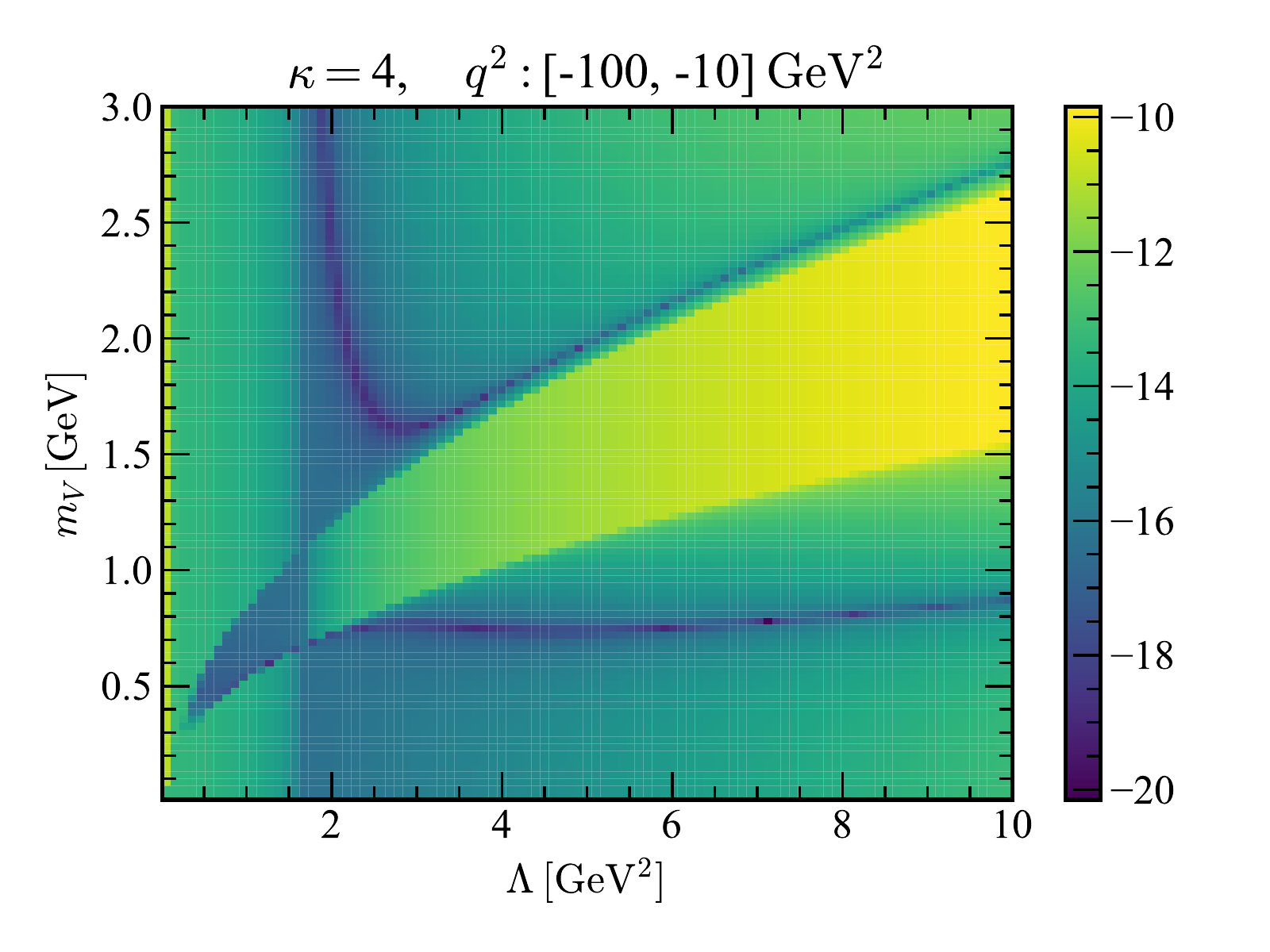}
\includegraphics[scale=0.35]{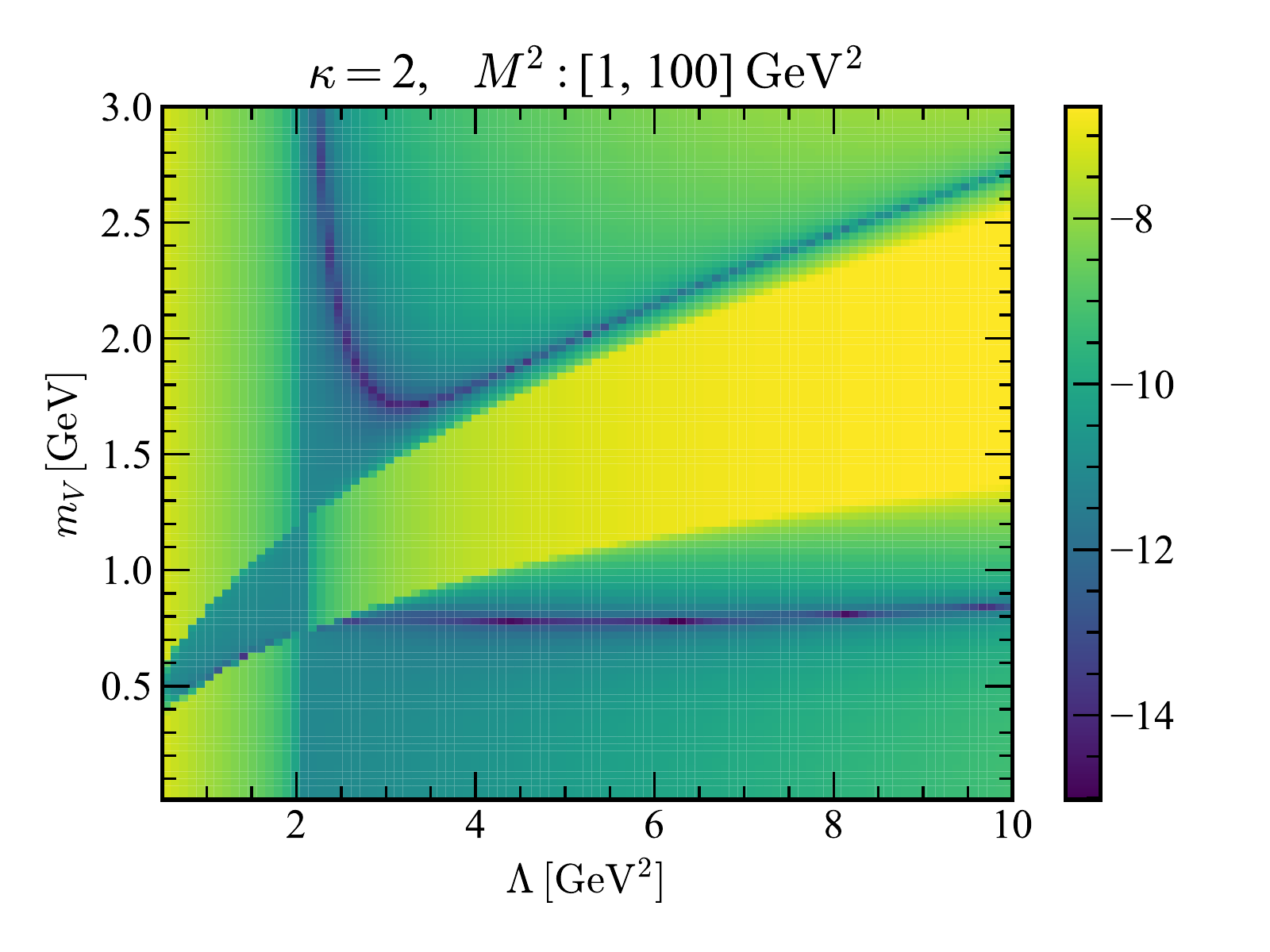}
\includegraphics[scale=0.35]{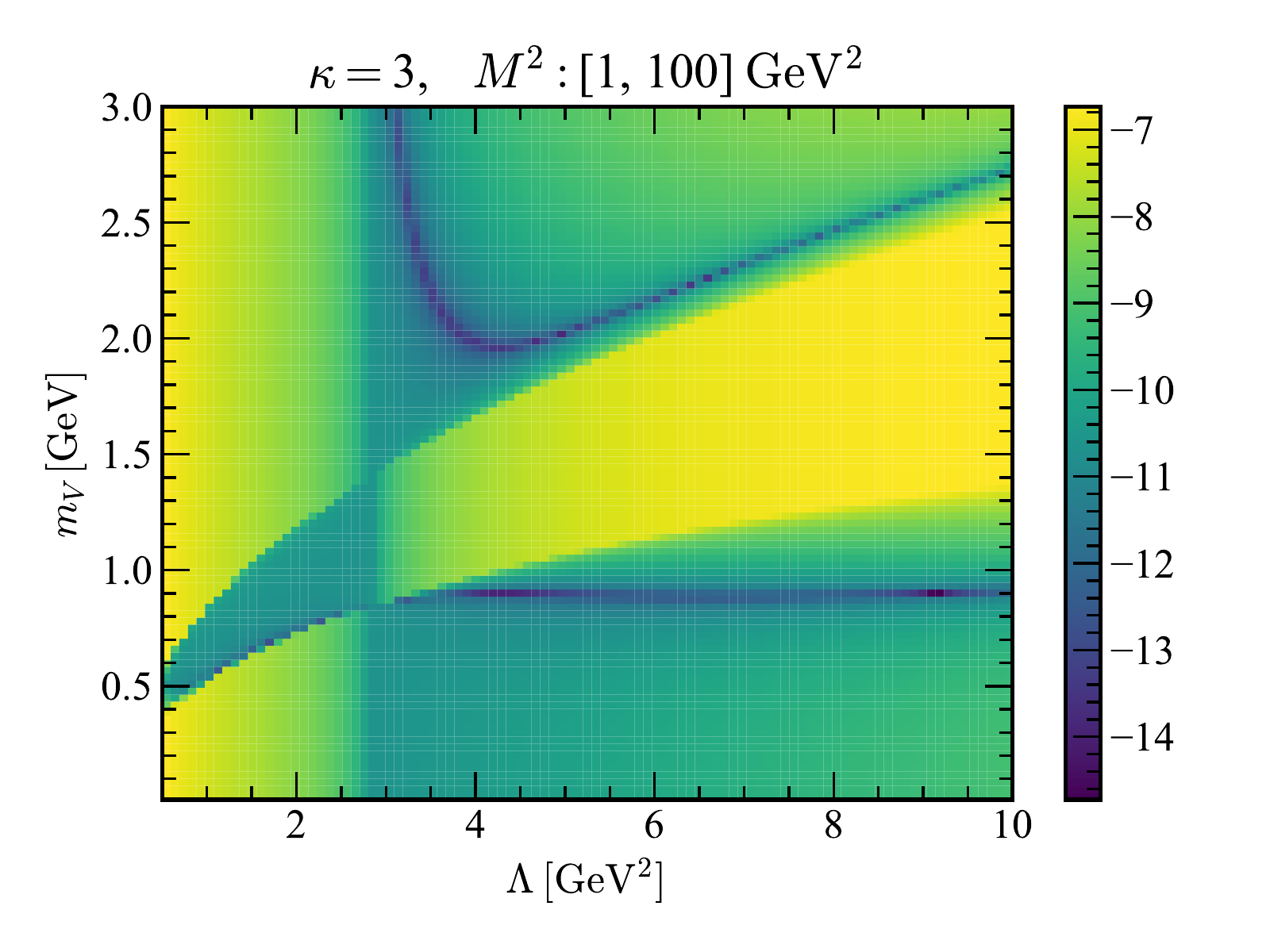}
\includegraphics[scale=0.35]{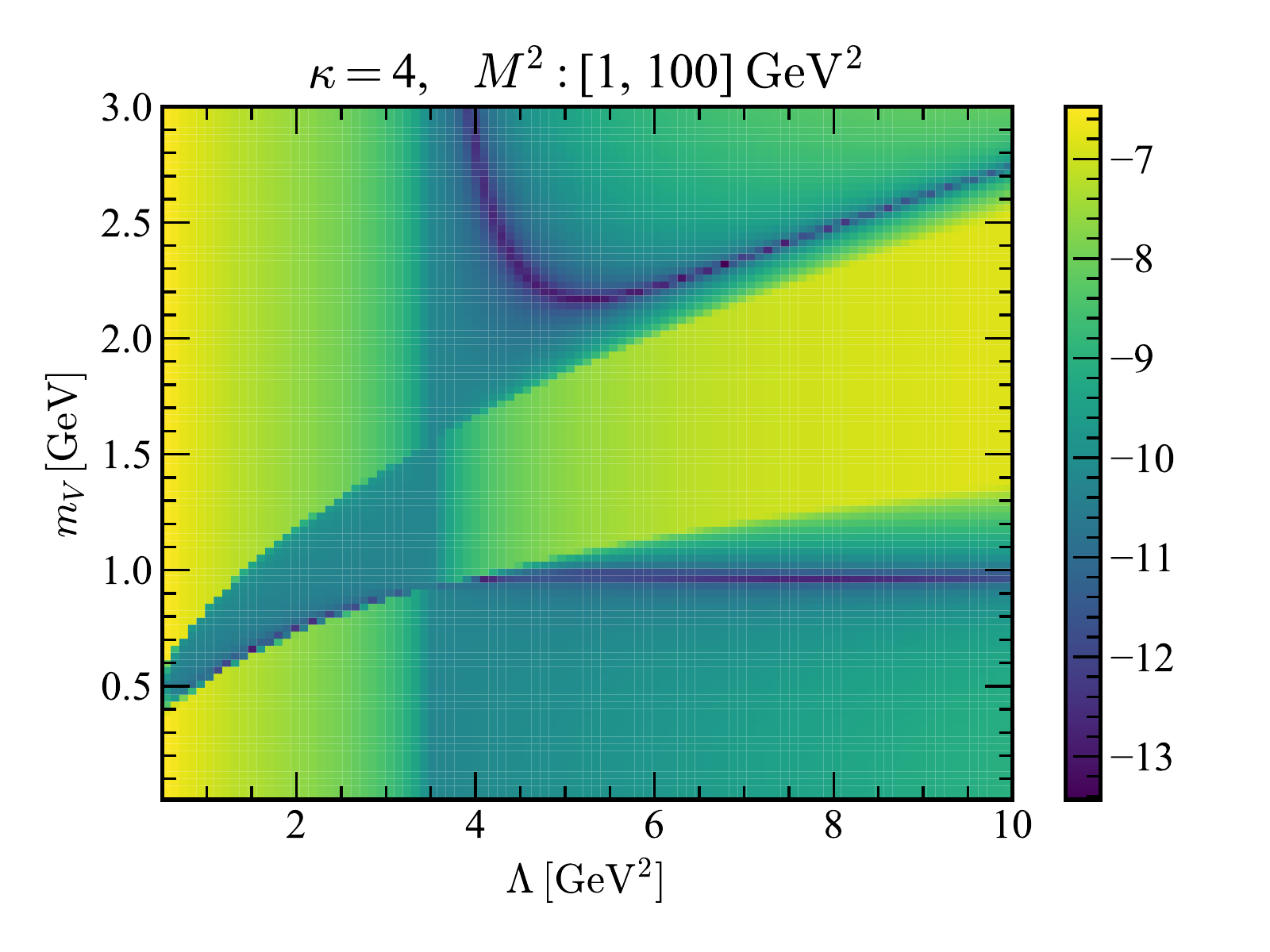}
\caption{\label{fig1}
Minimum distributions of RSS defined in Eq.~(\ref{dev}) on the $\Lambda$-$m_V$ planes
for $\kappa=2$, 3, and 4 with the input ranges $(-100\;{\rm GeV}^2,-1\;{\rm GeV}^2)$ in
$q^2$, $(-100\;{\rm GeV}^2,-10\;{\rm GeV}^2)$ in $q^2$, and
$(1\;{\rm GeV}^2,100\;{\rm GeV}^2)$ in $M^2$.}
\end{figure}

\begin{figure}
\includegraphics[scale=0.35]{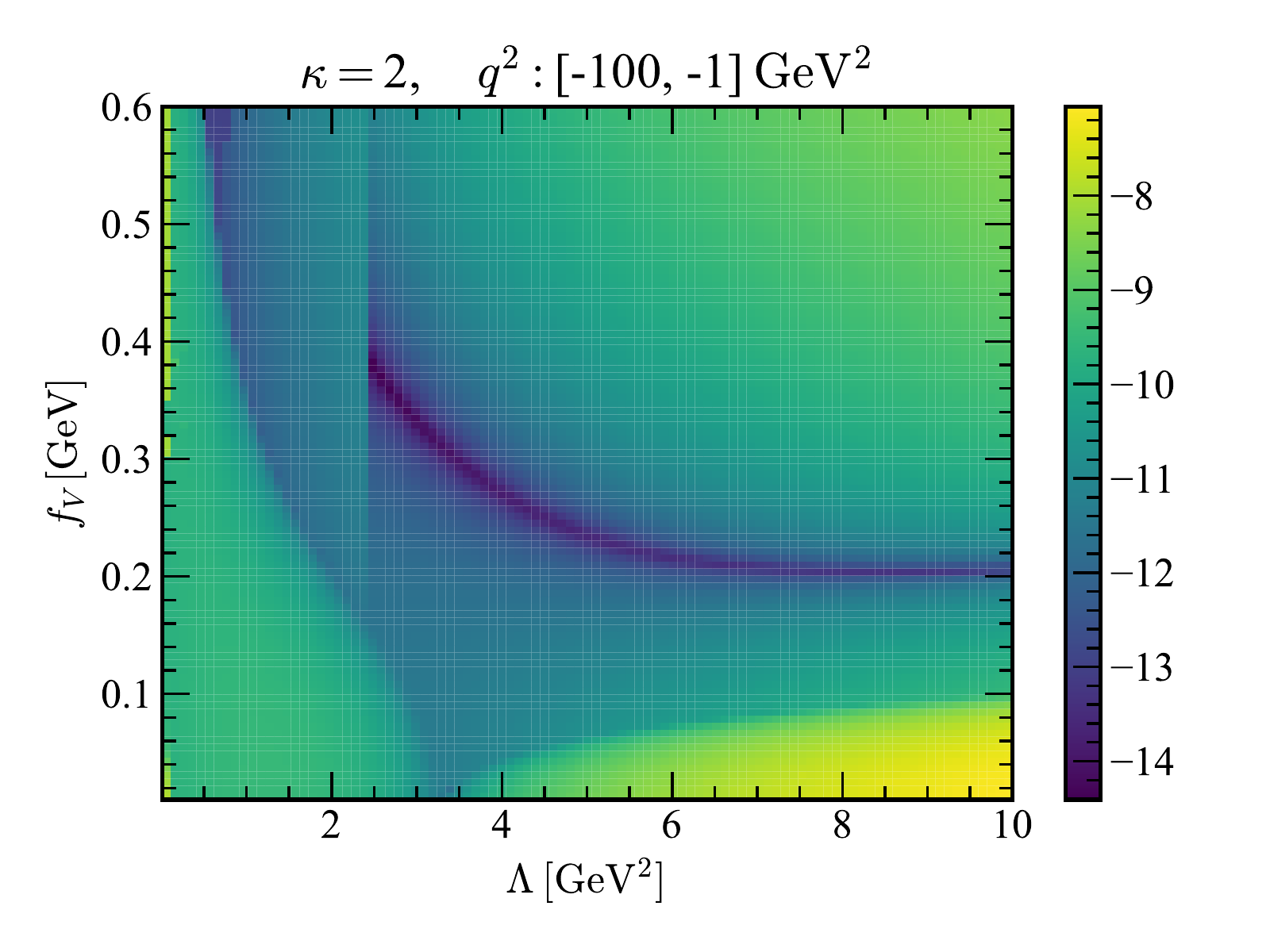}
\includegraphics[scale=0.35]{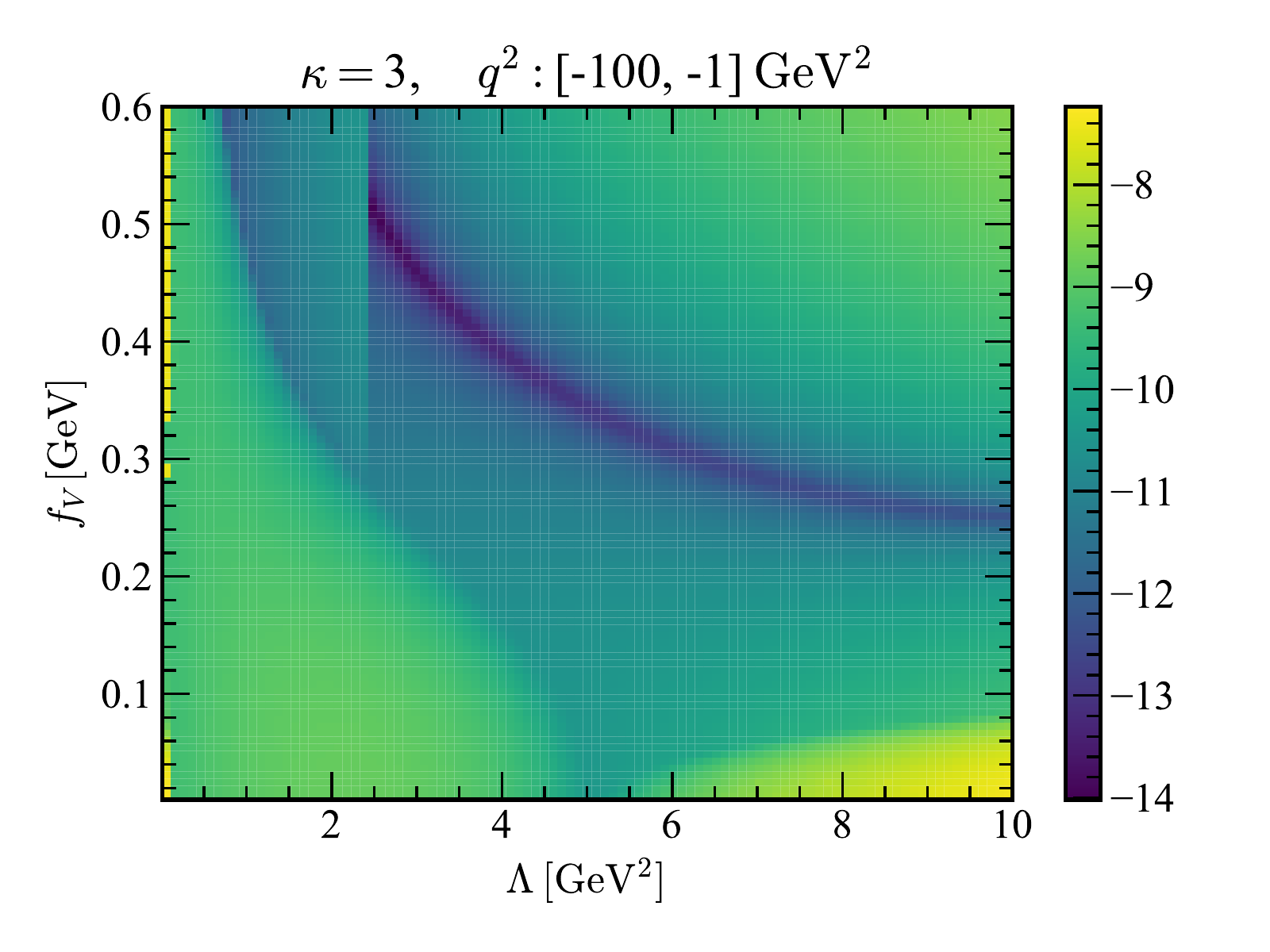}
\includegraphics[scale=0.35]{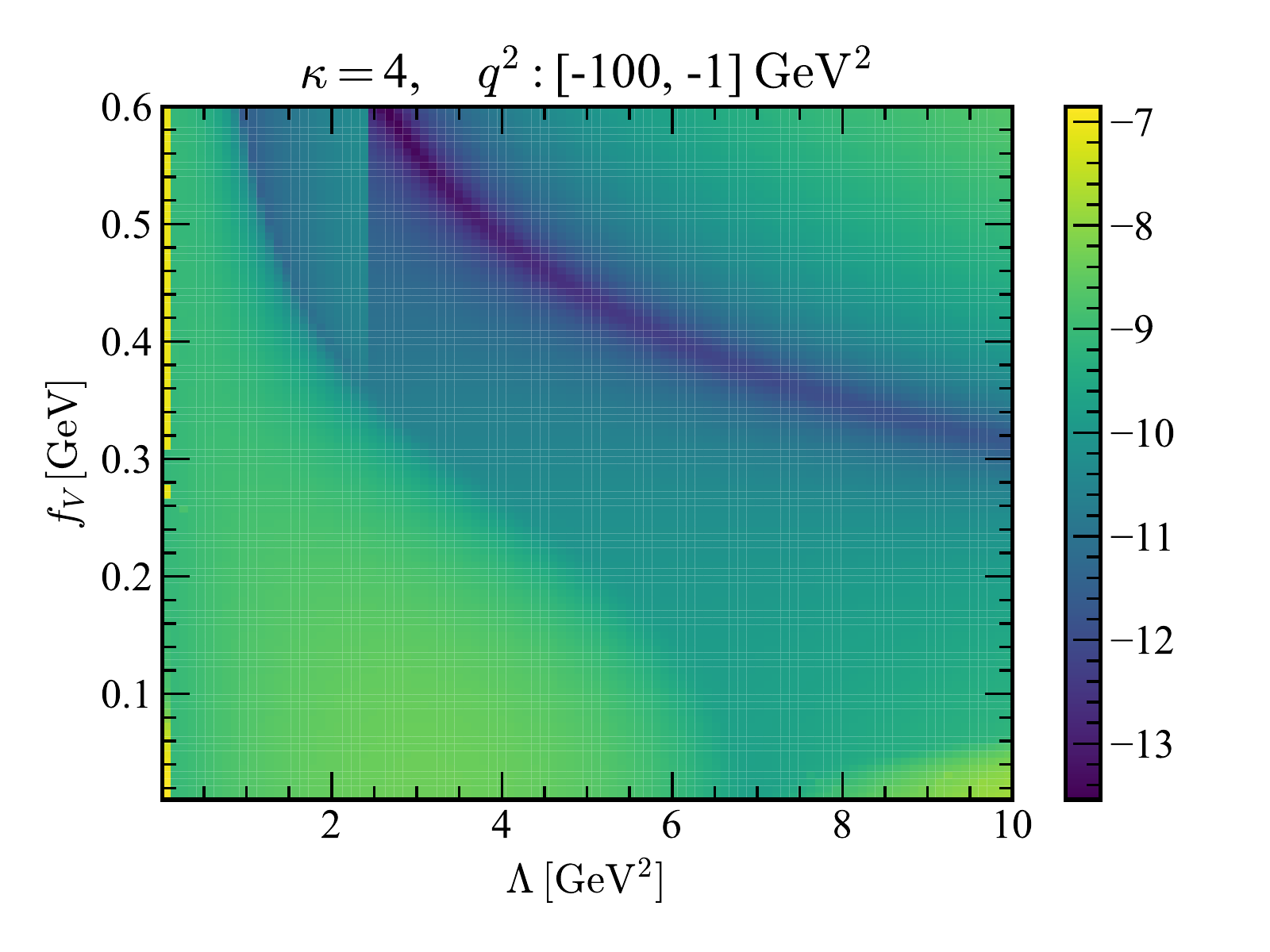}
\includegraphics[scale=0.35]{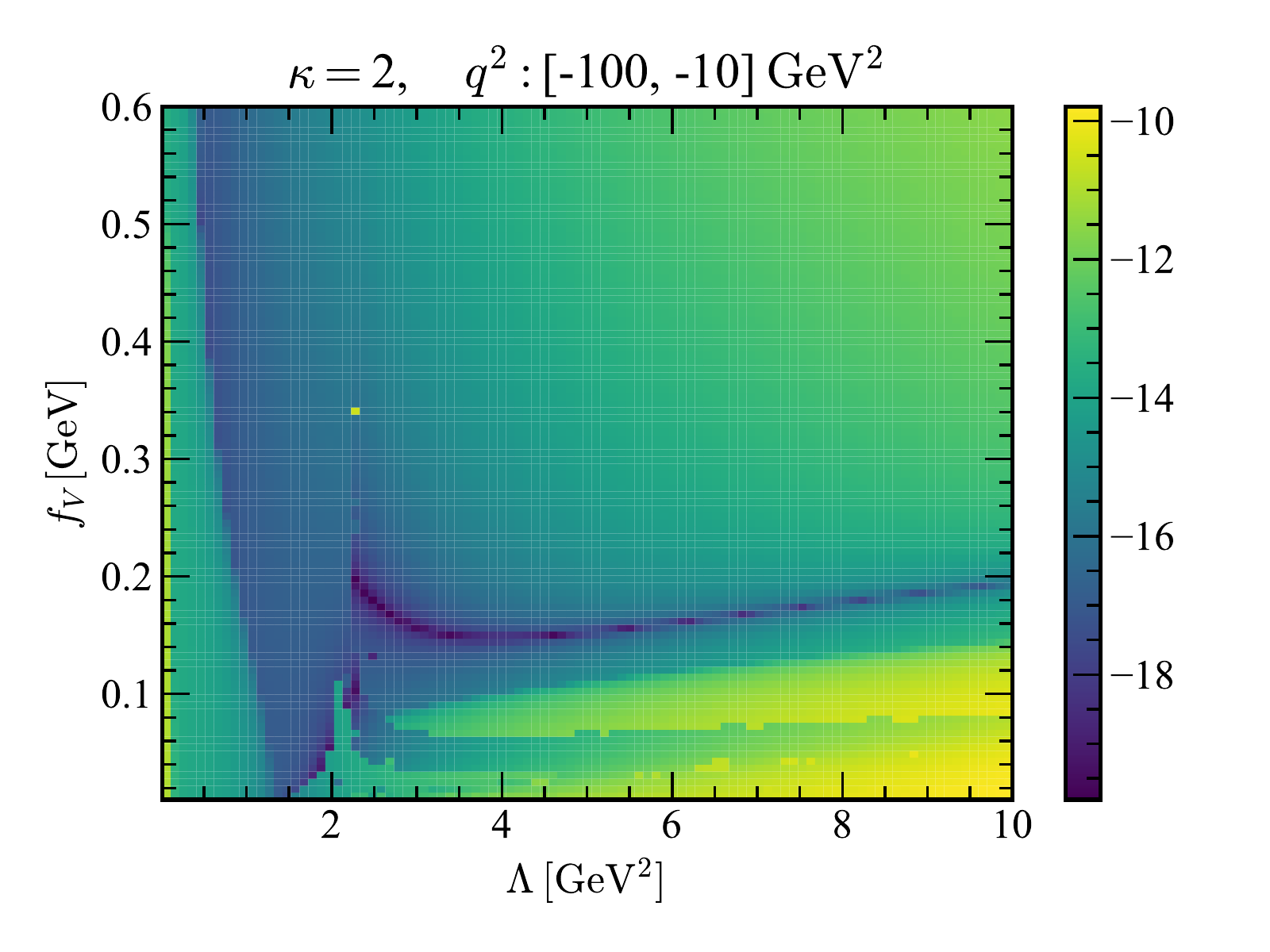}
\includegraphics[scale=0.35]{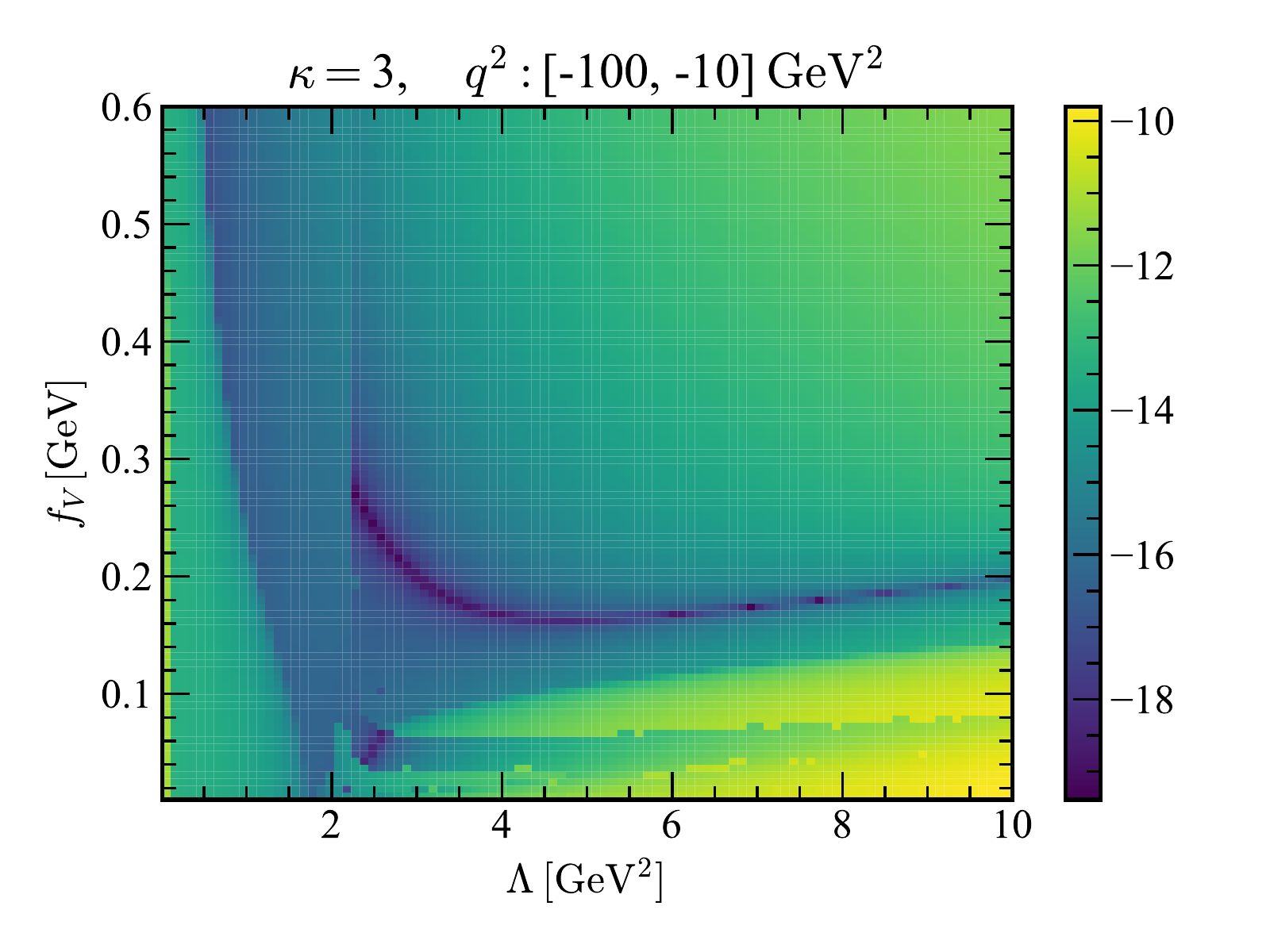}
\includegraphics[scale=0.35]{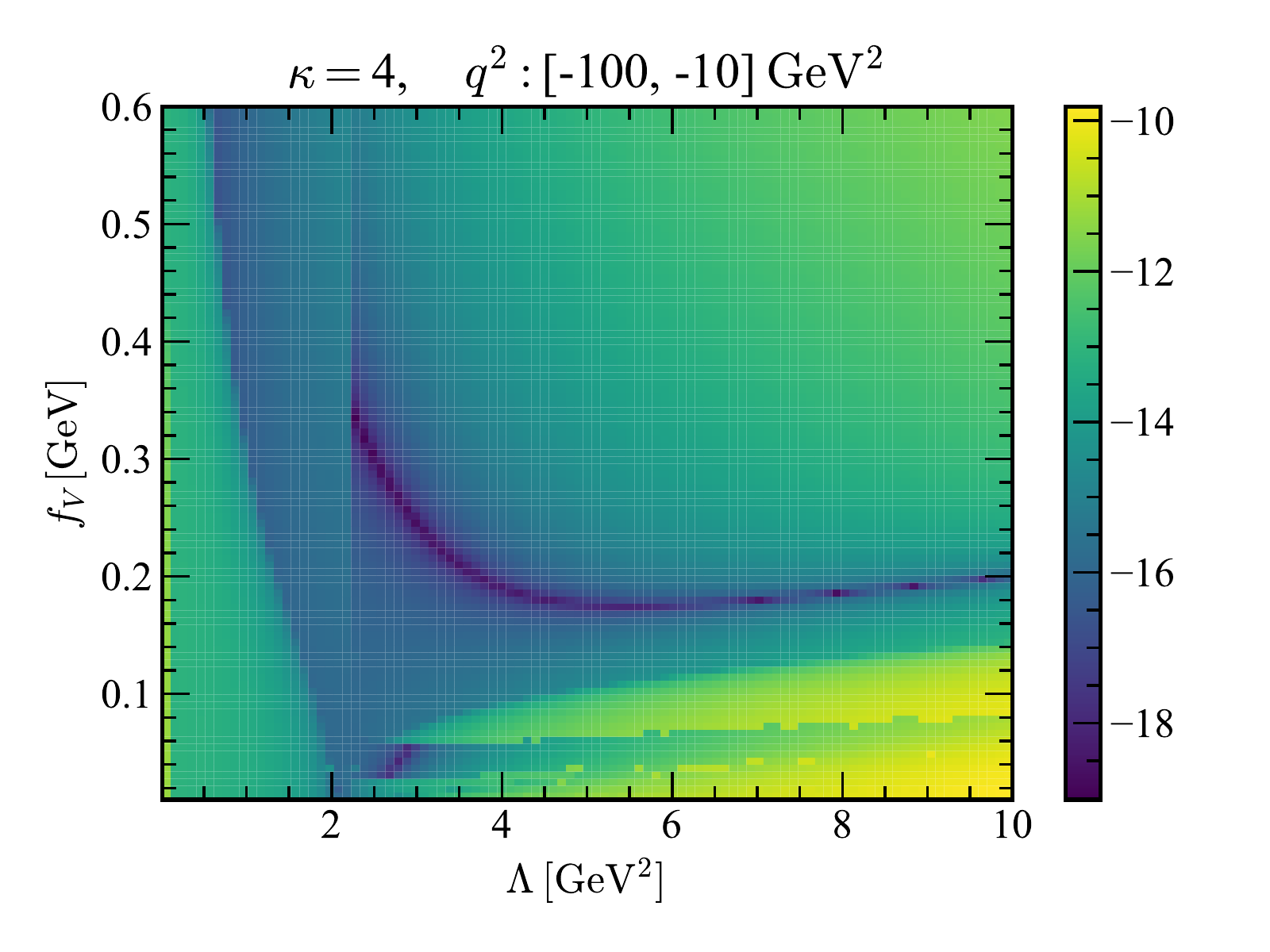}
\includegraphics[scale=0.35]{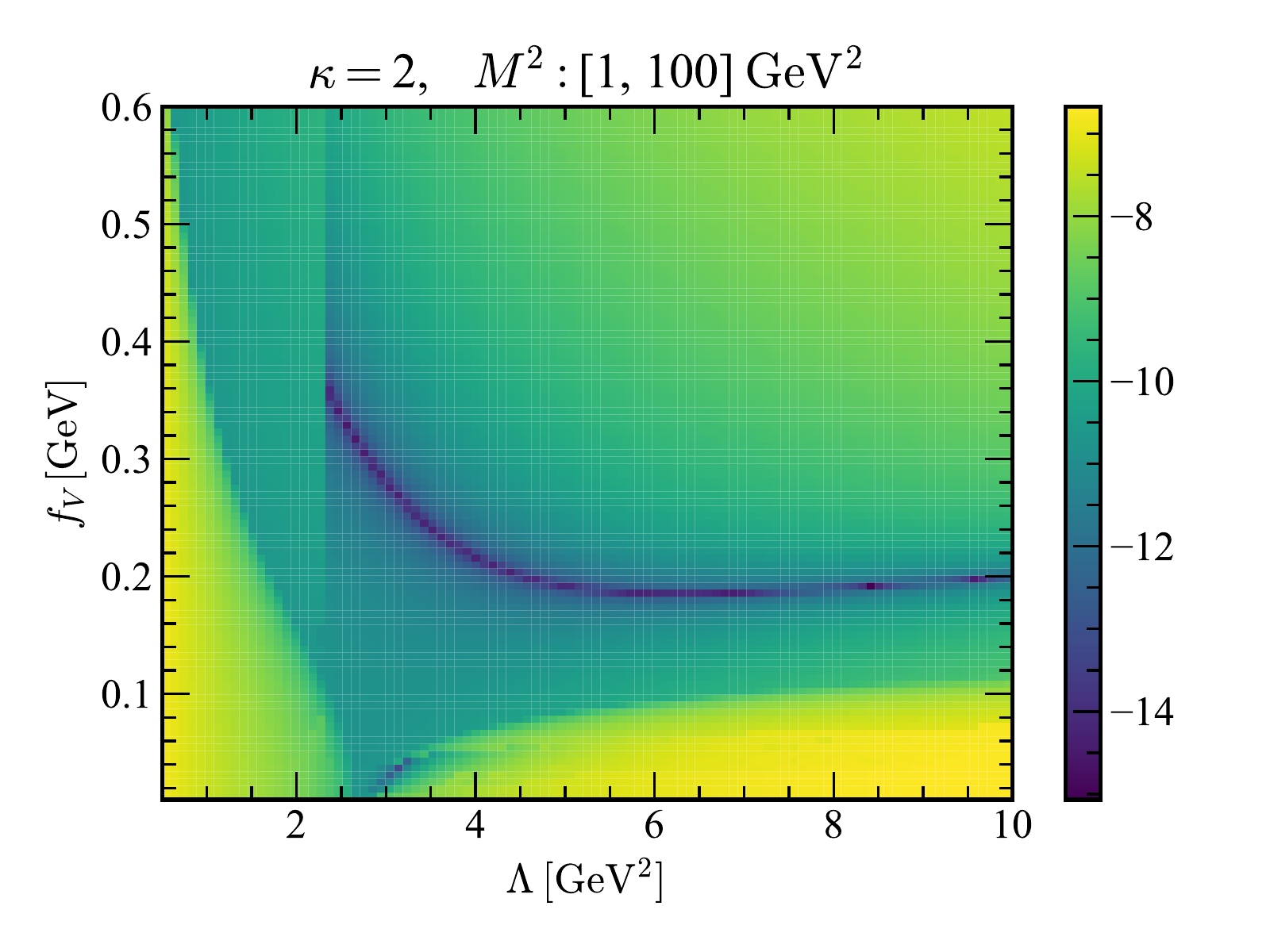}
\includegraphics[scale=0.35]{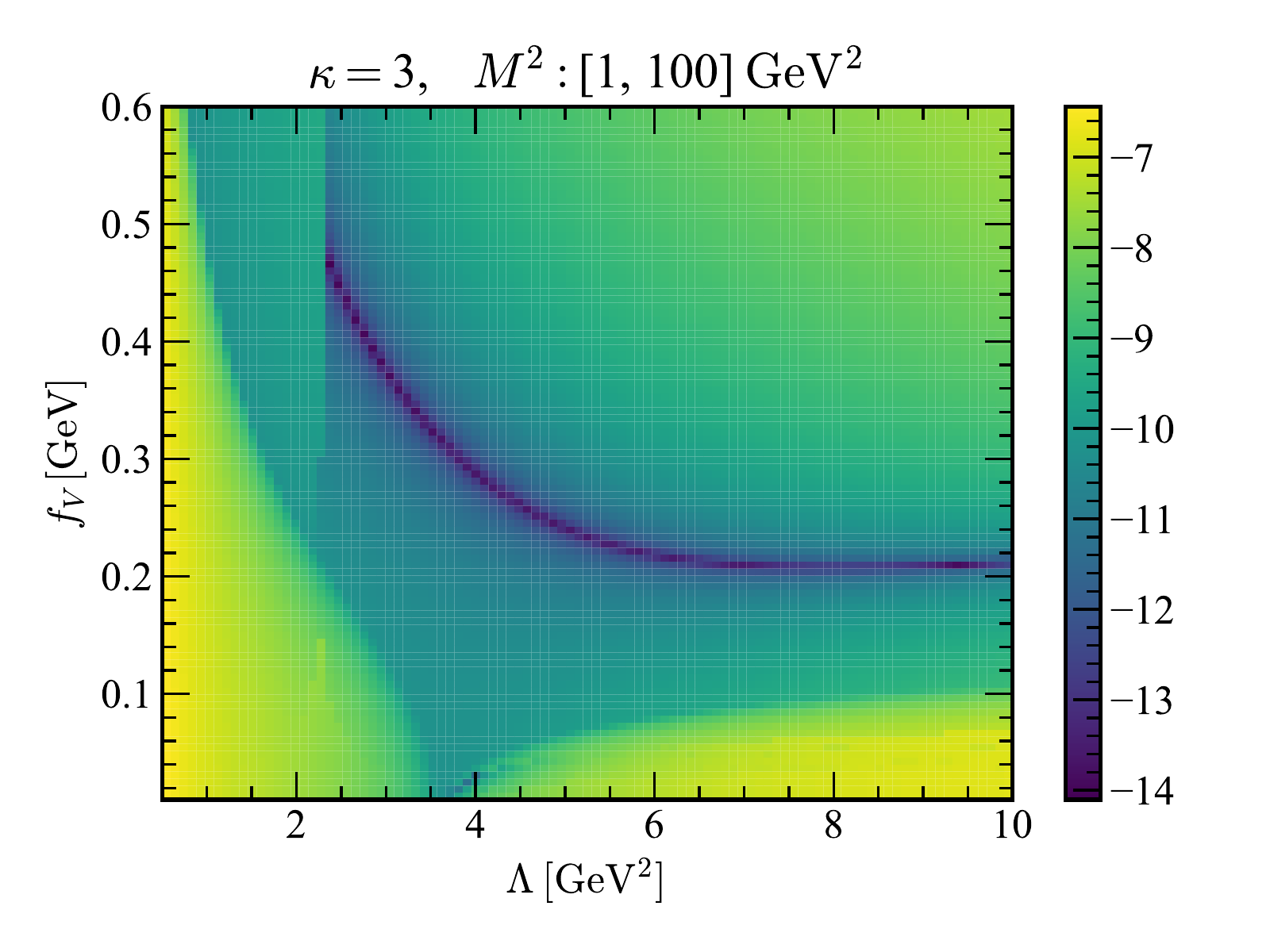}
\includegraphics[scale=0.35]{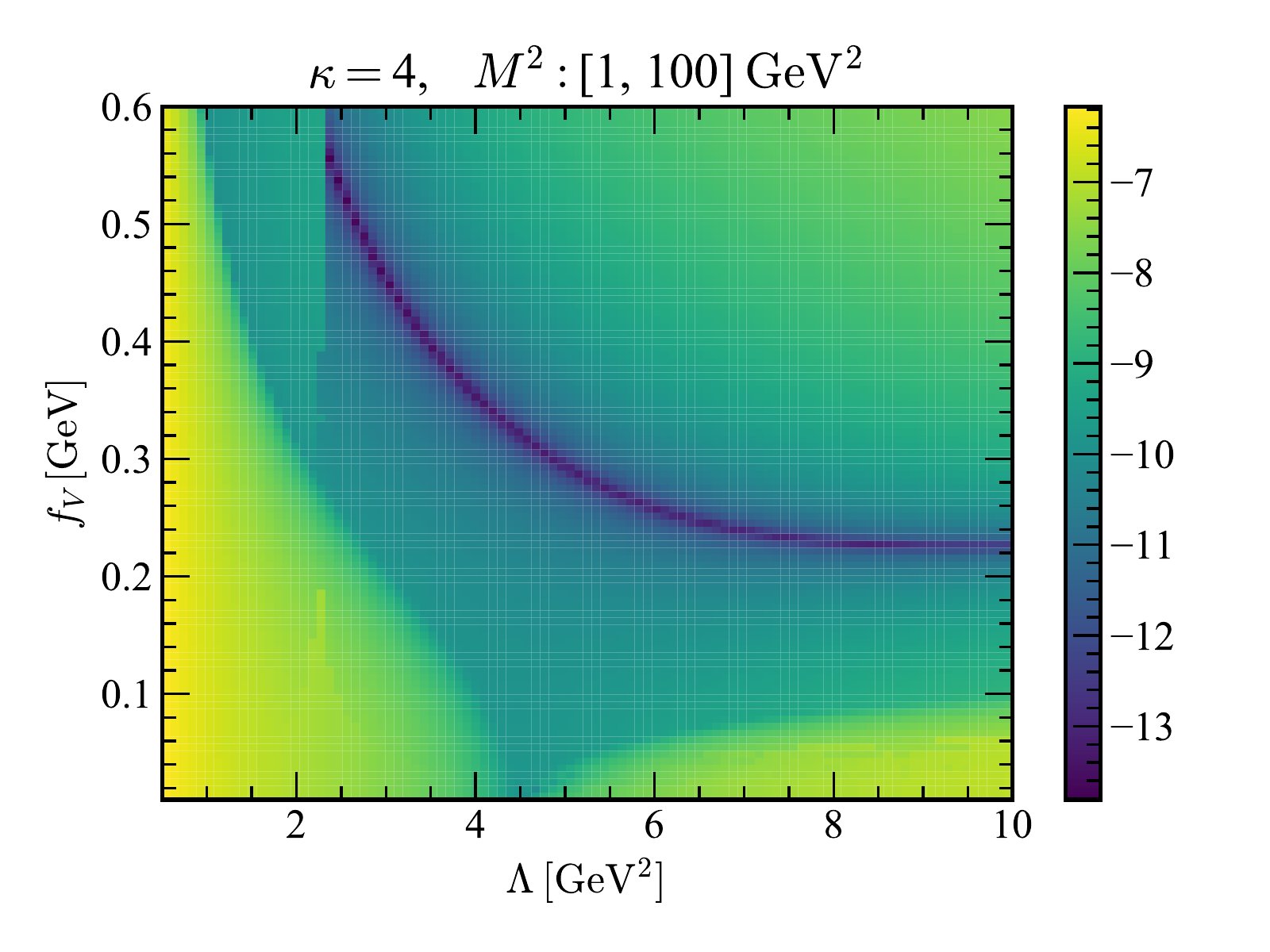}
\caption{\label{fig2}
Minimum distributions of RSS defined in Eq.~(\ref{dev}) on the $\Lambda$-$f_V$ planes
for $\kappa=2$, 3, and 4 with the input ranges $(-100\;{\rm GeV}^2,-1\;{\rm GeV}^2)$ in
$q^2$, $(-100\;{\rm GeV}^2,-10\;{\rm GeV}^2)$ in $q^2$, and
$(1\;{\rm GeV}^2,100\;{\rm GeV}^2)$ in $M^2$.}
\end{figure}

The scanning over all the free parameters reveals the minima of the RSS defined in Eq.~(\ref{dev}). 
We present the distributions of the RSS minima on the $\Lambda$-$m_V$ planes in 
Fig.~\ref{fig1} and on the $\Lambda$-$f_V$ planes in Fig.~\ref{fig2}, where each array contains 
three columns of plots for $\kappa=2$, 3, and 4,
and three rows for the OPE inputs from the ranges $(-100\;{\rm GeV}^2,-1\;{\rm GeV}^2)$ in $q^2$, 
$(-100\;{\rm GeV}^2,-10\;{\rm GeV}^2)$ in $q^2$, and $(1\;{\rm GeV}^2,100\;{\rm GeV}^2)$ in $M^2$. 
The reason why the input point is extended to $q^2=-100\;{\rm GeV}^2$ ($M^2=100\;{\rm GeV}^2$) 
is that we intend to find a solution to Eq.~(\ref{sum1}) (Eq.~(\ref{ivb})) in a large range of
$q^2$ ($M^2$). Nontrivial landscapes of the RSS minima 
are observed, which imply the resonance masses $m_V$ and the decay constants $f_V$ preferred by 
the sum rule in Eq.~(\ref{sum1}) or (\ref{ivb}). A point on a curve of deep color, 
having RSS about $10^{-14}$ ($10^{-18}$) relative to $10^{-8}$ ($10^{-10}$) from outside the curve
in the first and third (second) rows, represents an approximate solution to the sum rules.
A solution in the segment of the curve with deeper color is closer to the exact
solution, and the finite length of this segment hints the existence of multiple solutions. A value of 
$\Lambda$ labels the scale, below which the nonperturbative continuum contribution starts 
to deviate from the OPE input, so its variation affects the solutions of $m_V$ and $f_V$. This 
explains the dependence of the preferred $m_V$ and $f_V$ on $\Lambda$, described by the 
minimum distributions. 

It is expected that the power corrections would be enhanced with the input range 
$(-100\;{\rm GeV}^2,-1\;{\rm GeV}^2)$ in $q^2$ compared to 
$(-100\;{\rm GeV}^2,-10\;{\rm GeV}^2)$, because the former covers the low $Q^2$ region.
The enhancement is reflected by the sensitivity of the RSS minimum distributions to the variation 
of $\kappa$ in the first row of plots stronger than in the second row. The dependence 
of the minimum distributions on $\kappa$ is also more obvious in the third row with
the input from low $M^2$. Note that the dimension-six four-quark condensate correction 
becomes comparable to the dimension-four gluon condensate correction, both being of order 
of $10^{-3}$, at $Q^2$ and $M^2$ as low as $O(1)$ GeV$^2$. 
The minimum distributions obtained from Eq.~(\ref{ivb}) 
are similar to those from Eq.~(\ref{sum1}): the minimum locations on the planes in the third rows 
are somewhat between those in the first and second rows of Figs.~\ref{fig1} and \ref{fig2}. 
It is understood, since the coefficient of the dimension-six condensate in Eq.~(\ref{ivb}) is  
half of that in Eq.~(\ref{sum1}), and this reduction can be mimicked by selecting
the input from a larger $Q^2$ region for Eq.~(\ref{sum1}).
The above similarity supports the equivalence of Eqs.~(\ref{sum1}) and (\ref{ivb}), and 
our postulation that the Borel transformation is not needed, once sum rules are treated as an 
inverse problem. We will focus only on Eq.~(\ref{sum1}) for numerical analyses from now on.
As to the input range, we pick up the one, where the perturbative term is relatively 
more important than the condensate corrections, and the OPE is sufficiently convergent,
namely, $(-100\;{\rm GeV}^2,-10\;{\rm GeV}^2)$ in $q^2$. It has been checked that the input range 
$(-100\;{\rm GeV}^2,-5\;{\rm GeV}^2)$ in $q^2$ leads to the 
minimum distributions on the $\Lambda$-$m_V$ and $\Lambda$-$f_V$ planes almost
identical to the middle rows of Figs.~\ref{fig1} and \ref{fig2}, respectively.

It is interesting to see that two branches of minimum distributions appear on the $\Lambda$-$m_V$
planes with a gap between them, and the lower ones, being roughly flat (independent of $\Lambda$), 
are located around $m_V\approx 0.8$ GeV, which is close to the $\rho(770)$ meson mass $m_{\rho(770)}$. 
It has been claimed, based on a stable analytic extrapolation \cite{Buchert:1992zj}, that
the present perturbative amplitude in the deep Euclidean region produces a prominent bump 
structure in the resonance region. Here we have explicitly shown that the ground state $\rho(770)$
is predicted by our formalism. The depth of color in the second row of Fig.~\ref{fig1} indicates 
that global minima along the lower distributions appear in the range 2 GeV$^2<\Lambda< 4$ GeV$^2$. 
We point out that the minimum distribution in the central plot of Fig.~\ref{fig1} is 
flat, up to $\Lambda=6$ GeV$^2$, a behavior which can be regarded as kind of stability.
The predicted $\rho(770)$ meson mass read off from the above range is not sensitive to the parameter 
$\kappa$: it increases by about 10\% when $\kappa$ changes from 3 to 4, consistent with what was 
observed in \cite{Wang:2016sdt}. The upper minimum distributions in the $\Lambda$-$m_V$ plots 
imply that the single pole parametrization in Eq.~(\ref{imp1}) allows a larger $m_V$ to be a 
solution to Eq.~(\ref{sum1}). Note that Eq.~(\ref{imp1}) simply parametrizes the contribution from
a resonance and a continuum, so the resonance does not correspond to the $\rho(770)$ meson a priori.
It is likely that the contribution from an excited state and a continuum also obeys the Fredholm 
equation, and serves as one of the allowed multiple solutions. Therefore,
we conjecture that the upper distributions are associated with excited 
states, whose significance will be explored in the next subsection. 

There is only a single RSS minimum distribution on each $\Lambda$-$f_V$ plane in Fig.~\ref{fig2}. 
It is possible, if the ground state and the first excited state had similar decay constants. 
The plots in the second row imply a weaker dependence of the minimum distributions on $\Lambda$
than in the first and third rows. These minimum distributions
are located around $f_V\approx 0.2 $ GeV, close to the decay constant of the $\rho(770)$ 
meson. A larger $\kappa$ value, ie., a larger four-quark condensate
tends to increase $f_V$. In particular, the second row of Fig.~\ref{fig2} reveals global minima in the range 
2 GeV$^2<\Lambda< 4$ GeV$^2$, the same as in the second row of Fig.~\ref{fig1}. This consistency hints 
that the best solutions to the sum rule in Eq.~(\ref{sum1}) can accommodate the physical values
of the $\rho(770)$ meson mass and decay constant simultaneously.

\begin{figure}
\includegraphics[scale=0.40]{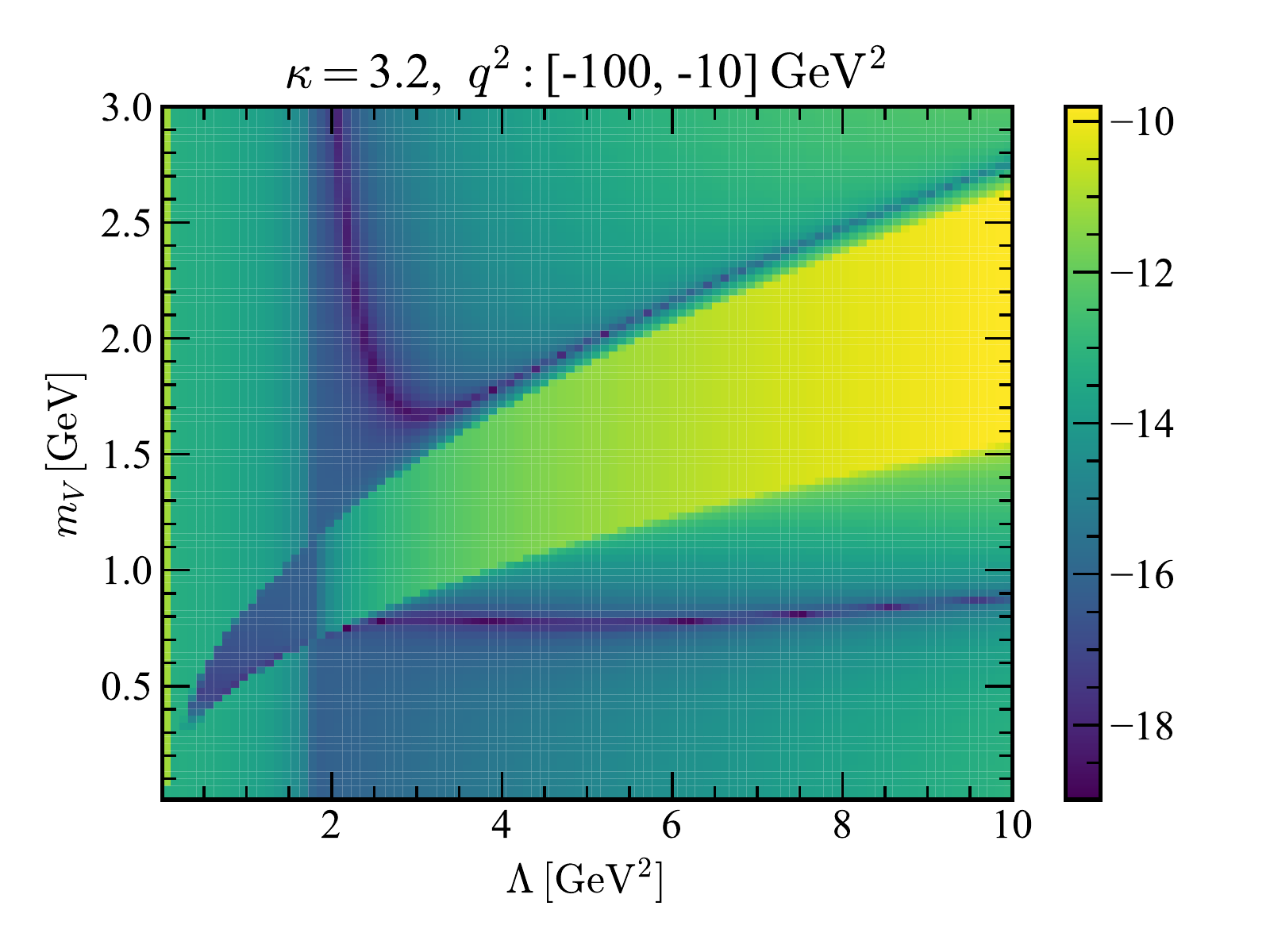}
\includegraphics[scale=0.40]{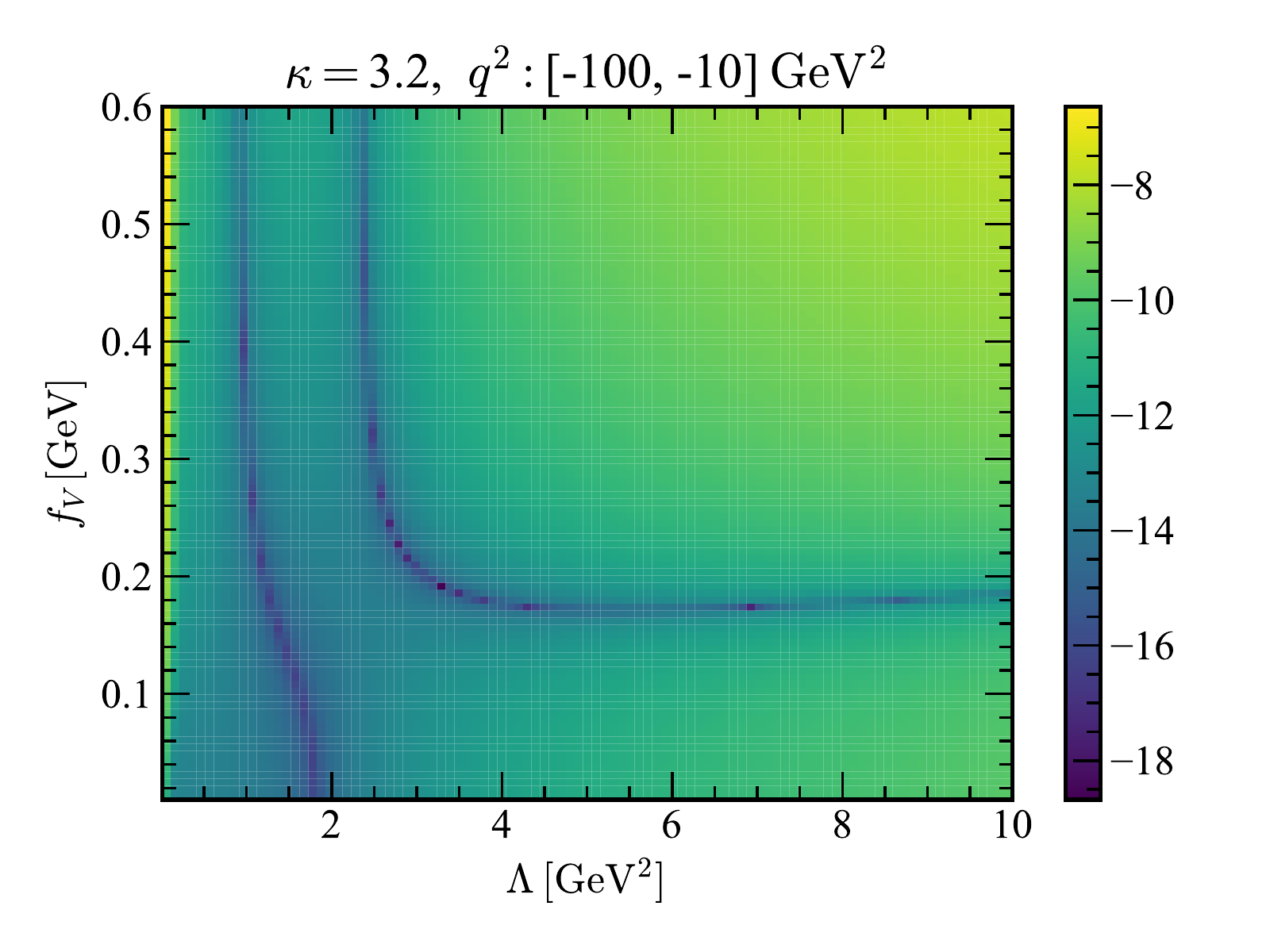}

(a)\hspace{6cm}(b)
\caption{\label{fig3}
Minimum distributions of RSS (a) on the $\Lambda$-$m_V$ plane
for $\kappa=3.2$ with the input range $(-100\;{\rm GeV}^2,-10\;{\rm GeV}^2)$ in
$q^2$, and (b) on the $\Lambda$-$f_V$ with $m_V$ being further set to $m_{\rho(770)}\approx 0.78$ GeV.}
\end{figure}

We first fix the factorization violation parameter $\kappa$ associated with the 
four-quark condensate using the $\rho(770)$ meson mass $m_{\rho(770)}\approx 0.78$ GeV, and adopt 
this $\kappa$ value for further analyses. It is straightforward to find that the $\rho(770)$ meson 
mass can be produced with $\kappa=3.2$, a value also preferred by \cite{SN09},
along the lower minimum distribution in a wide range 2 GeV$^2 <\Lambda < 6$ GeV$^2$ as shown 
in Fig.~\ref{fig3}(a). Since we have ensured that both the best fitted $m_V$ and
$f_V$ occur roughly in the same range of $\Lambda$, we determine $f_V$ in a less ambiguous
way by setting $m_V$ to $m_{\rho(770)}\approx 0.78$ GeV to avoid possible influence from
excited states. The resultant $\Lambda$-$f_V$ plot
from Eq.~(\ref{sum1}) with the input range $(-100\;{\rm GeV}^2,-10\;{\rm GeV}^2)$ in $q^2$ 
is displayed in Fig.~\ref{fig3}(b).
For consistency, we search for the global minimum in the range 2 GeV$^2 <\Lambda < 6$ GeV$^2$,
and find that the one located at $\Lambda=2.8$ GeV$^2$ on the L-shape 
distribution gives the decay constant 
$f_{\rho(770)}= 0.22$ GeV. The separation scale $\Lambda=2.8$ GeV$^2$ is supposed to be large 
enough for justifying the replacement of ${\rm Im}\Pi(s)$ by ${\rm Im}\Pi^{\rm pert}(s)$ in 
the second term on the right hand side of Eq.~(\ref{di2}). The above results of $m_{\rho(770)}$ and 
$f_{\rho(770)}$ agree with those in \cite{Tanabashi:2018oca}, from the lattice calculation 
\cite{Sun:2018cdr}, from the Bethe-Salpeter equation \cite{BL,WW,BKM}, and from the light-front 
quark model \cite{CJ}.

\begin{figure}
\includegraphics[scale=0.40]{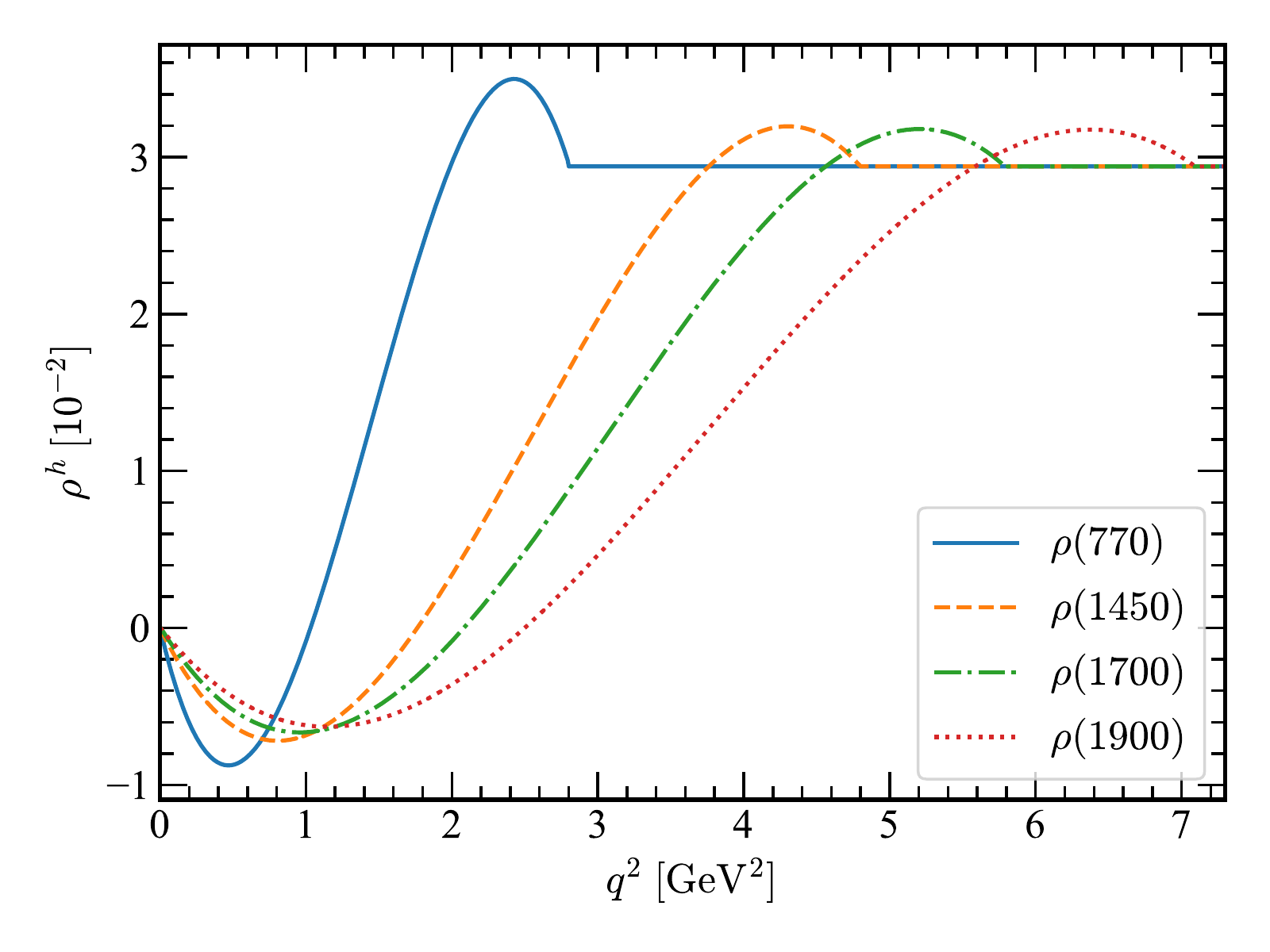}
\includegraphics[scale=0.40]{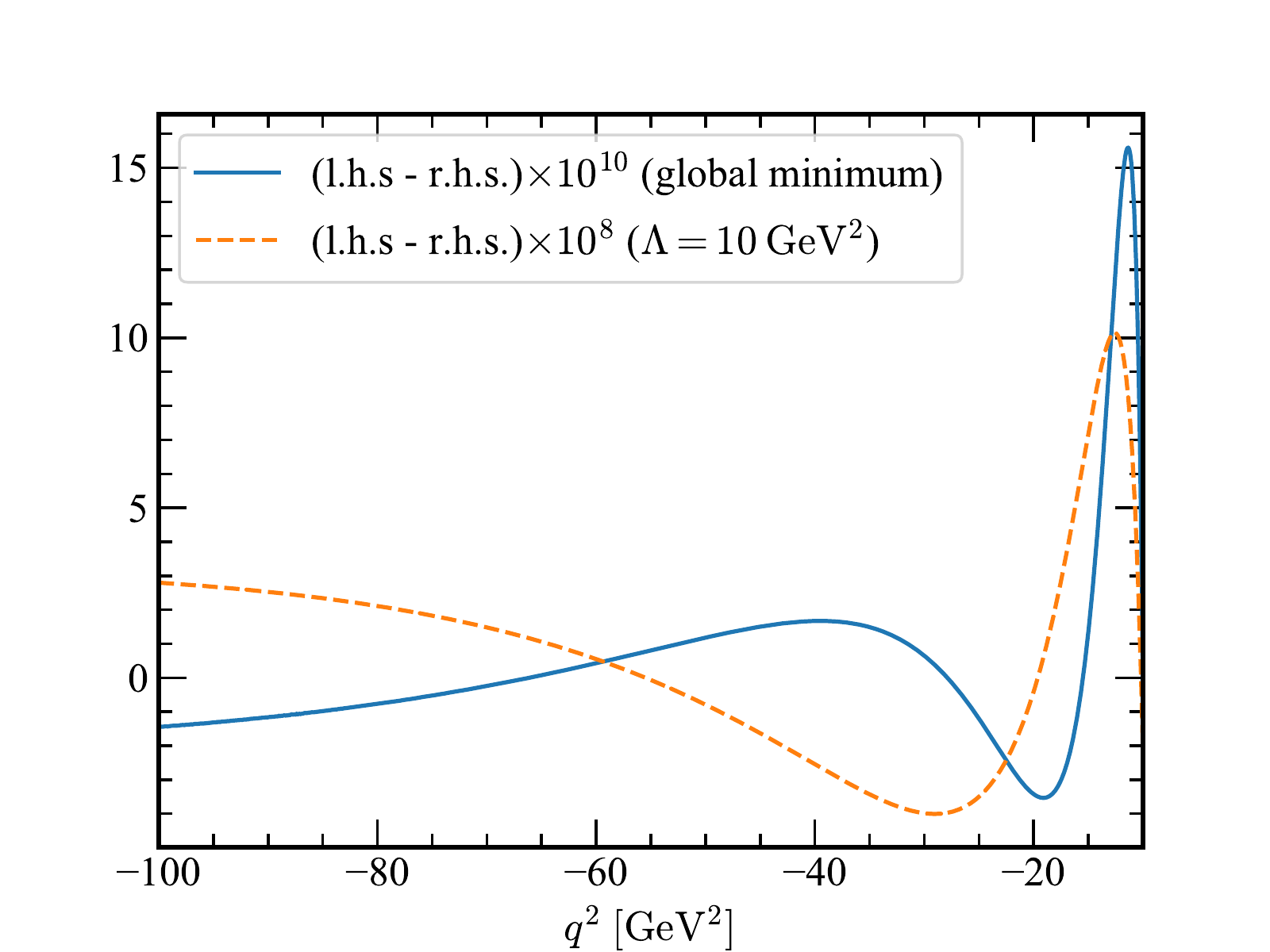}

(a)\hspace{6cm}(b)
\caption{\label{fig4}
(a) behavior of the spectral density function $\rho^h(q^2)$ in $q^2$ , and (b) difference between the two sides of
Eq.~(\ref{sum1}) for the best fit solution in 
Fig.~\ref{fig3}(b), and a solution on the minimum distribution located at $\Lambda=10$ GeV$^2$.}
\end{figure}

To test the sensitivity of our results to the OPE uncertainties, 
we vary the perturbative piece and the quark condensate $\langle \bar q q\rangle$
by $\pm 30\%$ separately. It is observed that our results are less sensitive to 
the variation of the former. Since the quark condensate
and the gluon condensate appear at the same power of $1/(q^2)^2$, the $30\%$
variation can include and mimic that from the gluon condensate. It is understood that 
the $\langle \bar q q\rangle^2$ term at the power $1/(q^2)^3$ also varies accordingly.
We find that the $\rho(770)$ meson mass, extracted
from the single-pole parametrization in Eq.~(\ref{sum1}), differs by only about $\pm 15\%$. 
It implies that our results in the present setup are stable, as the OPE uncertainties are taken into account.

We also read off the coefficients in the expansion of the spectral density function $\rho^h(q^2)$
in terms of the Legendre polynomials, which correspond to the selected global minimum located at 
$\Lambda=2.8$ GeV$^2$ in Fig.~\ref{fig3}(b):
\begin{eqnarray}
b_0= 0.0126, \;\;\;\;b_1=0.0276,\;\;\;\; b_2= 0.0022,\;\;\;\; b_3= -0.0128.\label{coe1}
\end{eqnarray}
As two more Legendre polynomials $P_4$ and $P_5$ are included in the expansion, the global minimum 
shifts to $\Lambda=3.1$ GeV$^2$ with the corresponding decay constant $f_{\rho(770)}= 0.23$ GeV, 
and we have
\begin{eqnarray}
b_0= 0.0120, \;\;\;\;b_1=0.0308,\;\;\;\; b_2= -0.0040,\;\;\;\; b_3= -0.0202,
\;\;\;\; b_4= 0.0068,\;\;\;\; b_5=  0.0042.\label{coe2}
\end{eqnarray}
The stability of the coefficients $b_0$,..., $b_3$ and the smallness of $b_4$ and $b_5$ 
verify that the expansion up to the polynomial $P_3$ is enough. 

The behavior of the spectral density function $\rho^h(q^2)$ 
in $q^2$ for Eq.~(\ref{coe1}) is depicted in Fig.~\ref{fig4}(a), which differs dramatically from 
the step function in Eq.~(\ref{non}) based on the local quark-hadron duality. Note that the slope 
of the solved $\rho^h(q^2)$ is discontinuous at 
$q^2=\Lambda=2.8$ GeV$^2$, where it transits to the perturbative spectra density function, because we 
have not yet imposed the continuity constraint to the slope. 
The exact spectral density should approach to the perturbative one at a sufficiently large $\Lambda$, 
such that a solution becomes less sensitive to the choice of the separation scale $\Lambda$.
This explains the flatness of the minimum distribution for $\Lambda > 4$ GeV$^2$ in Fig.~\ref{fig3}(b).
The function $\rho^h(q^2)$ is slightly negative at $q^2$, where the pole is located. 
This negative contribution is expected to be compensated by that from the resonance, when its finite
width is taken into account. We mention that
the minimum distribution on the $\Lambda$-$f_V$ plane in Fig.~\ref{fig3}(b) becomes nearly vertical
at $\Lambda\approx 2.3$ GeV$^2$, corresponding to solutions with large $f_V$. It is easy to find that 
the corresponding function $\rho^h(q^2)$ is significantly negative at the $\rho(770)$ pole, so
$f_V$ must be large to compensate this negative contribution.

We compare the $q^2$ dependencies of the left hand side from the best fit solution and of the 
right hand side of Eq.~(\ref{sum1}) by showing their difference in Fig.~\ref{fig4}(b).  
For the purpose of comparison, we display a solution on the minimum distribution located at 
$\Lambda=10$ GeV$^2$ in Fig.~\ref{fig3}(b) with the corresponding decay constant $f_V= 0.18$ GeV, 
which is far away from the global minimum. It is obvious that the two sides of Eq.~(\ref{sum1}) match each 
other well in the former case, and that the difference between the two sides is about 100 times 
larger in the latter case.

\begin{figure}
\includegraphics[scale=0.40]{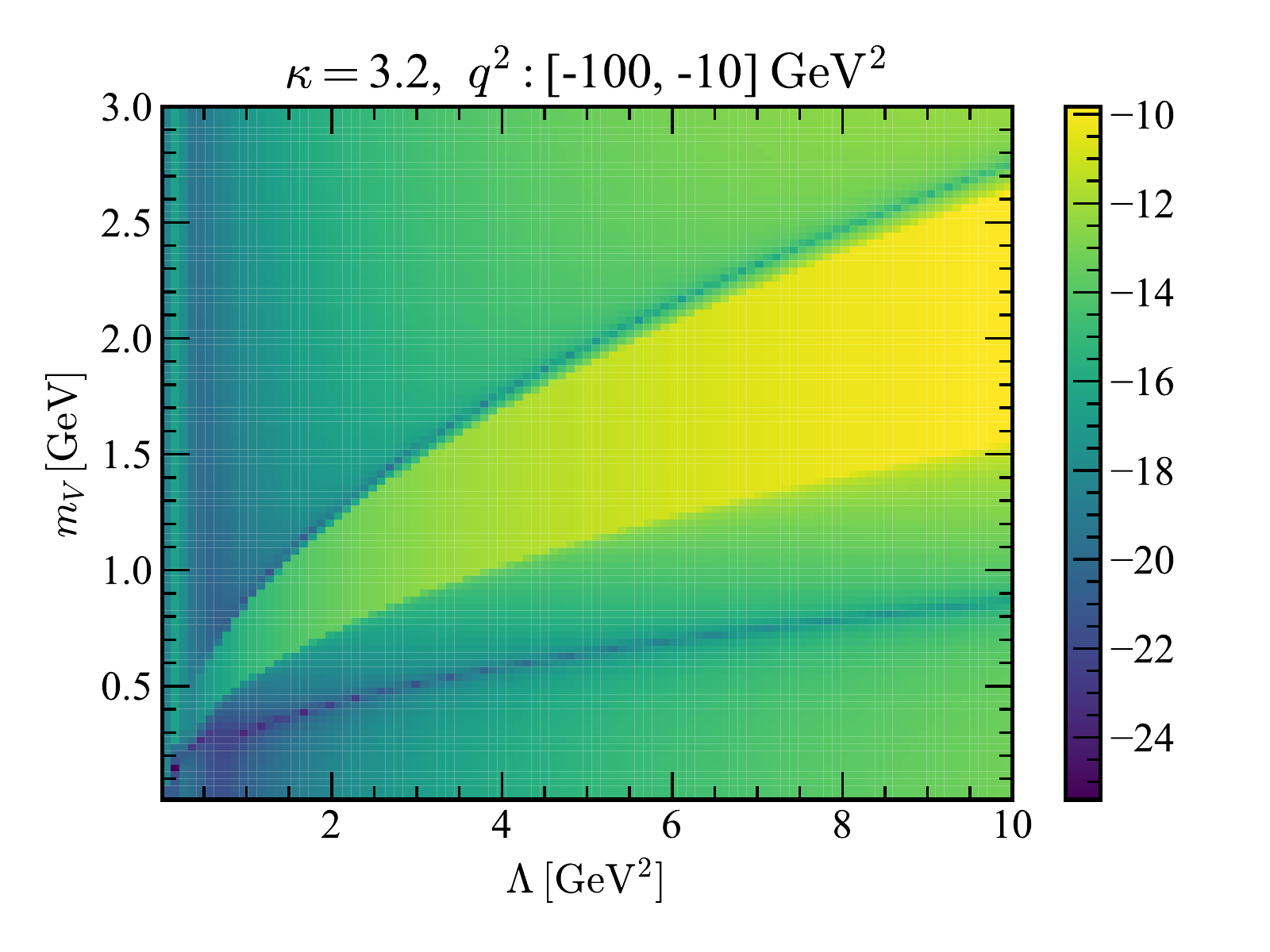}
\includegraphics[scale=0.40]{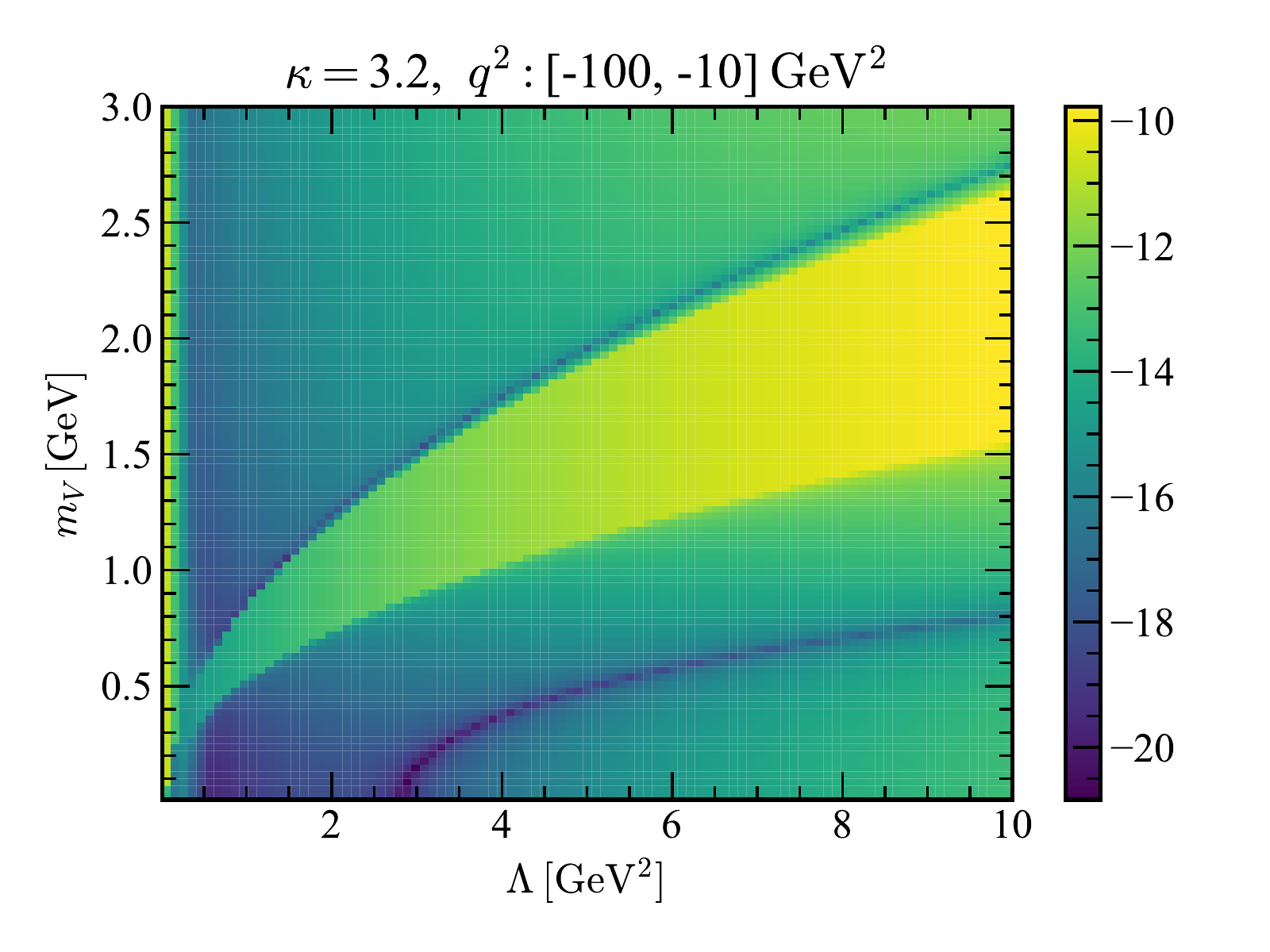}

(a)\hspace{6cm}(b)
\caption{\label{fig5}
Minimum distributions of RSS on the $\Lambda$-$m_V$ planes
for $\kappa=3.2$ with the input of (a) the perturbative piece only, and (b)  
without the $1/(q^2)^3$ power correction from the range $(-100\;{\rm GeV}^2,-10\;{\rm GeV}^2)$ in $q^2$.}
\end{figure}

Next we investigate how each term in the OPE input influences the emergence 
of the $\rho$ resonances in Fig.~\ref{fig1}. The RSS minimum distributions on the $\Lambda$-$m_V$ plane 
for the two cases, with only the perturbative piece and without the $1/(q^2)^3$ power correction, 
are presented in Figs.~\ref{fig5}(a) and \ref{fig5}(b), respectively. It is seen that both the minimum 
distributions grow with $\Lambda$ in the former without a stable region. 
It implies that the perturbative piece alone does not induce a bound state.
When the $1/(q^2)^2$ terms, ie., the gluon and two-quark condensates, are turned on, the 
lower minimum distribution in Fig.~\ref{fig5}(b) shifts toward the larger $\Lambda$ region, 
but still does not exhibit a stable value 
of $m_V$. The upper minimum distribution remains as dim as in  Fig.~\ref{fig5}(a).
When all the terms on the right hand sides of Eq.~(\ref{sum1}) are present, the stable ground
state mass appears, and the upper minimum distribution also gets enhanced as shown in Fig.~\ref{fig3}.
The observation is that all the terms in the OPE work together to generate the $\rho$ resonance, 
and the four-quark condensate is more crucial for its emergence.

\begin{figure}
\includegraphics[scale=0.40]{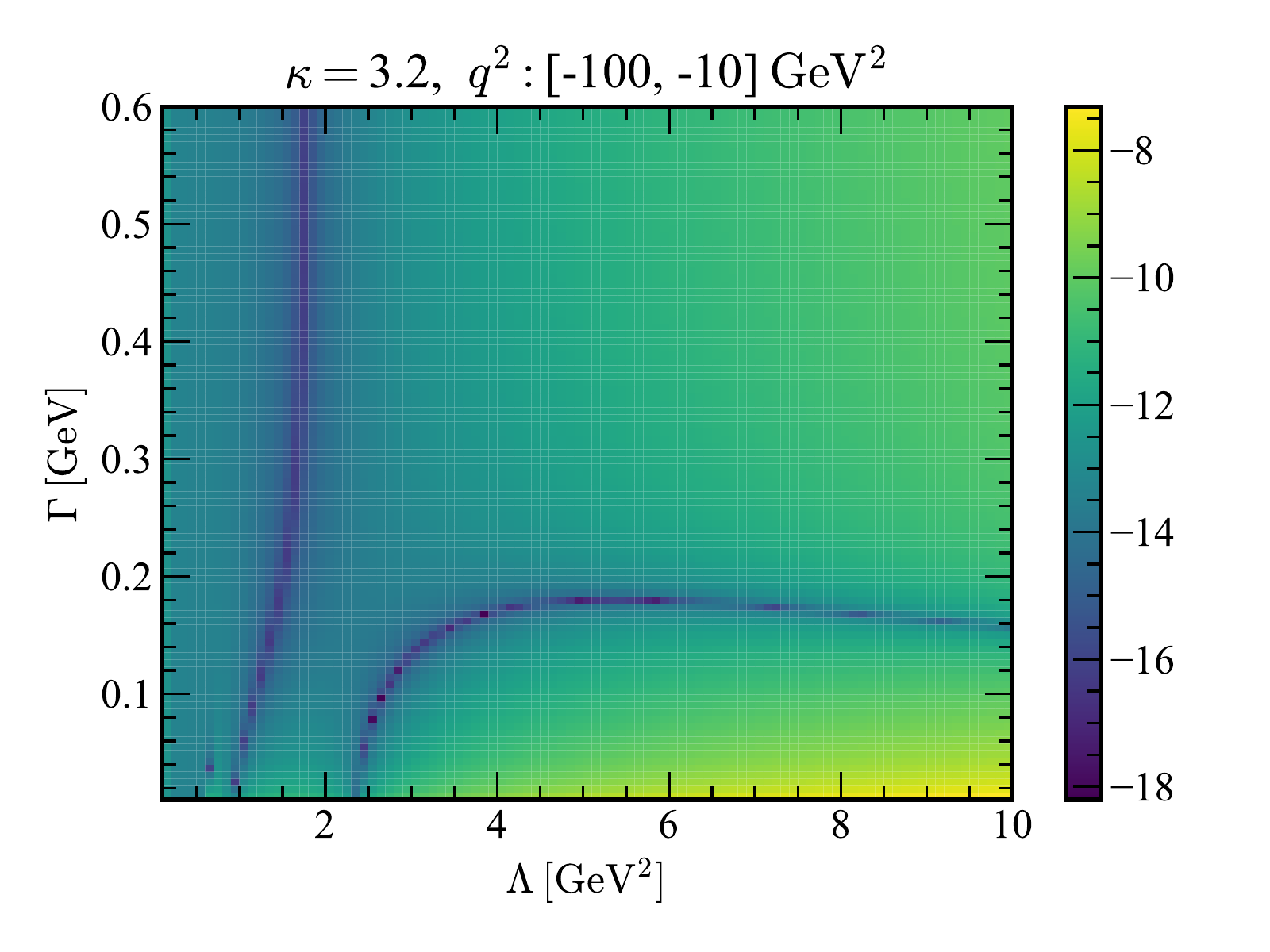}
\caption{\label{fig6}
Minimum distribution of RSS on the $\Lambda$-$\Gamma$ plane
for $\kappa=3.2$ with the input from the range $(-100\;{\rm GeV}^2,-10\;{\rm GeV}^2)$ in $q^2$.}
\end{figure}

Below we extract the $\rho$ meson decay width from our formalism by inserting the
$\pi^+\pi^-$ state and other multi-hadron states into the correlator in Eq.~(\ref{cur}), 
among which the matrix element $\langle 0|J_\mu|\pi^+\pi^-\rangle$ defines
the time-like pion form factor. The $\rho$ resonance contributes to the spectral
density dominantly, which is parametrized as \cite{DP,Wang:2016sdt}
\begin{eqnarray}
{\rm Im}\Pi(q^2)=\frac{1}{24\pi}\frac{m_V^4+m_V^2\Gamma^2}{(q^2-m_V^2)^2+m_V^2\Gamma^2}+\pi\rho^h(q^2),
\label{imbw}
\end{eqnarray}
with the width $\Gamma$.
The first term in the above expression corresponds to the time-like pion form factor,
which describes the decay of a $\rho$ meson, produced by the current $J_\mu$, into a pion pair. 
No three pion states are involved here, which arise from an $\omega$ resonance suppressed by the
isospin-1 current. The boundary condition $\rho^h(q^2=\Lambda)=a$ ought to be 
modified into ${\rm Im}\Pi(q^2=\Lambda)=a\pi$ due to the finite distribution
of the form factor in $q^2$. We set $m_V = m_{\rho(770)}$, 
and analyze the minimum distribution on the $\Lambda$-$\Gamma$ plane, which
is exhibited in Fig.~\ref{fig6}. Other parametrizations for the time-like pion form
factor, such as the Breit-Wigner one in \cite{Fischer:2020fvl}, have been tested, and similar minimum 
distributions are obtained. It is noticed that the global minimum located at
$\Lambda=4.3$ GeV$^2$ gives $\Gamma=0.17$ GeV, close to the value in \cite{Tanabashi:2018oca}.
We mention that it is difficult to reproduce the width of the $\rho(770)$ meson with any 
reasonable precision in the Bayesian approach \cite{Gubler:2010cf}, because of the insufficient 
sensitivity of the detailed $\rho(770)$ peak form to the OPE input.

\subsection{Excited States}

The upper minimum distributions on the $\Lambda$-$m_V$ planes in Fig.~\ref{fig1},
with a gap above the lower ones, hint strongly the existence of other 
resonances, though a single pole parametrization for the spectral density
was adopted in Eq.~(\ref{imp1}). Combining the implication of the $\Lambda$-$f_V$ plots in 
Fig.~\ref{fig2}, we speculate that an excited $\rho$ state with the decay constant 
around 0.2 GeV can also satisfy the sum rule in Eq.~(\ref{sum1}).
Compared to the lower minimum distributions, referred to the ground state $\rho(770)$, 
the minimum distributions associated with excited states exhibit more significant sensitivity 
to the variation of the separation scale $\Lambda$. This is understandable, because more excited states,
which are denser in the mass spectrum, will be covered as $\Lambda$ increases, such that a single
pole parametrization becomes less proper.
To study properties of an excited state, we modify the spectral density 
in Eq.~(\ref{imp1}) into
\begin{eqnarray}
{\rm Im}\Pi(q^2)=\pi f_{\rho(770)}^2\delta(q^2-m_{\rho(770)}^2)
+\pi f_V^2\delta(q^2-m_V^2)+\pi\rho^h(q^2),\label{imp2}
\end{eqnarray}
where the ground state mass and decay constant have been fixed to
$m_{\rho(770)}=0.78$ GeV and $f_{\rho(770)}=0.22$ GeV determined in the previous subsection.
The first excited state has been moved out of the continuum and treated 
as the second isolated resonance in the above parametrization.
Certainly, the RSS definition in Eq.~(\ref{dev}) is also modified accordingly 
with two resonance terms being included. 

\begin{figure}
\includegraphics[scale=0.40]{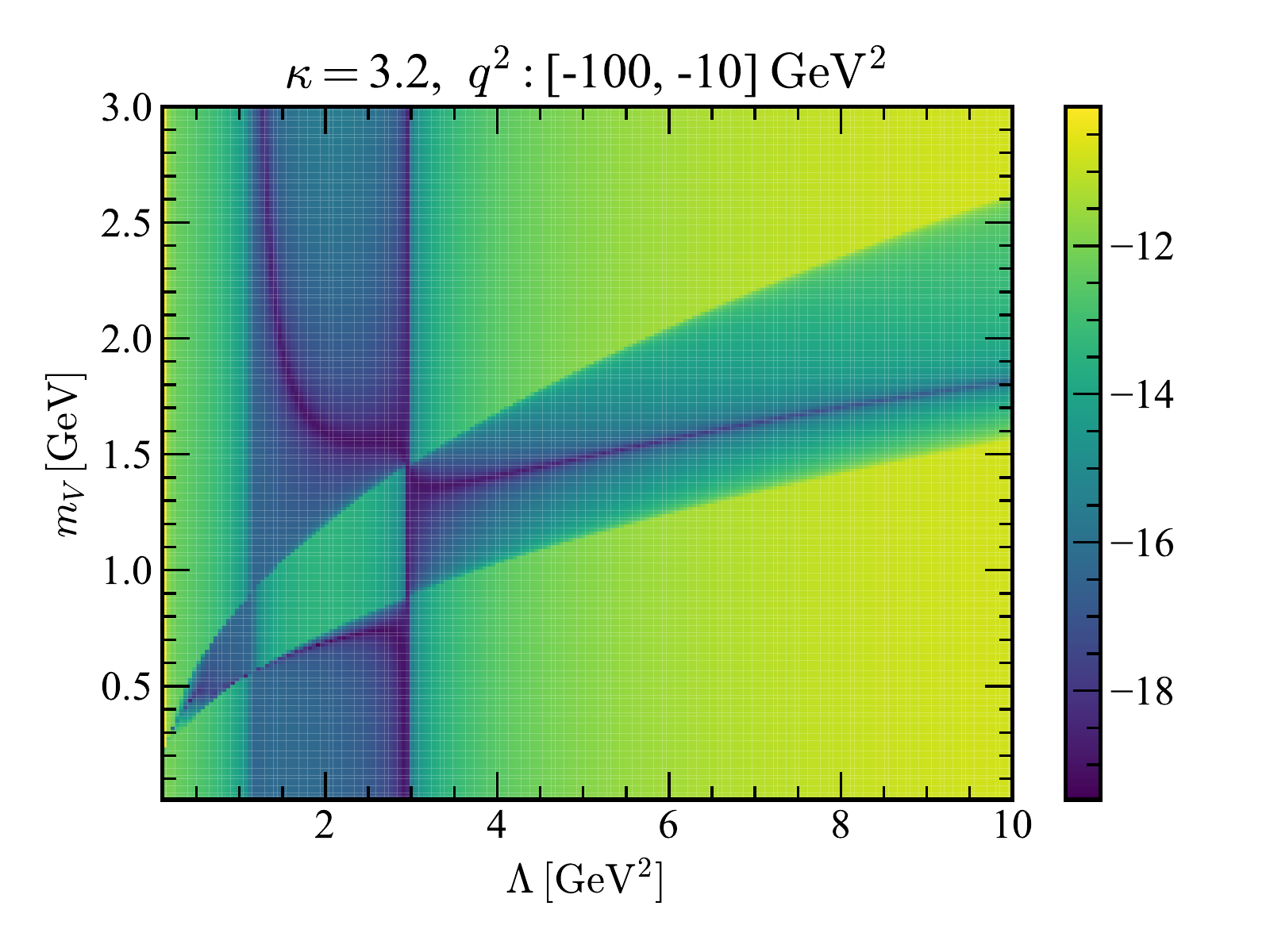}
\includegraphics[scale=0.40]{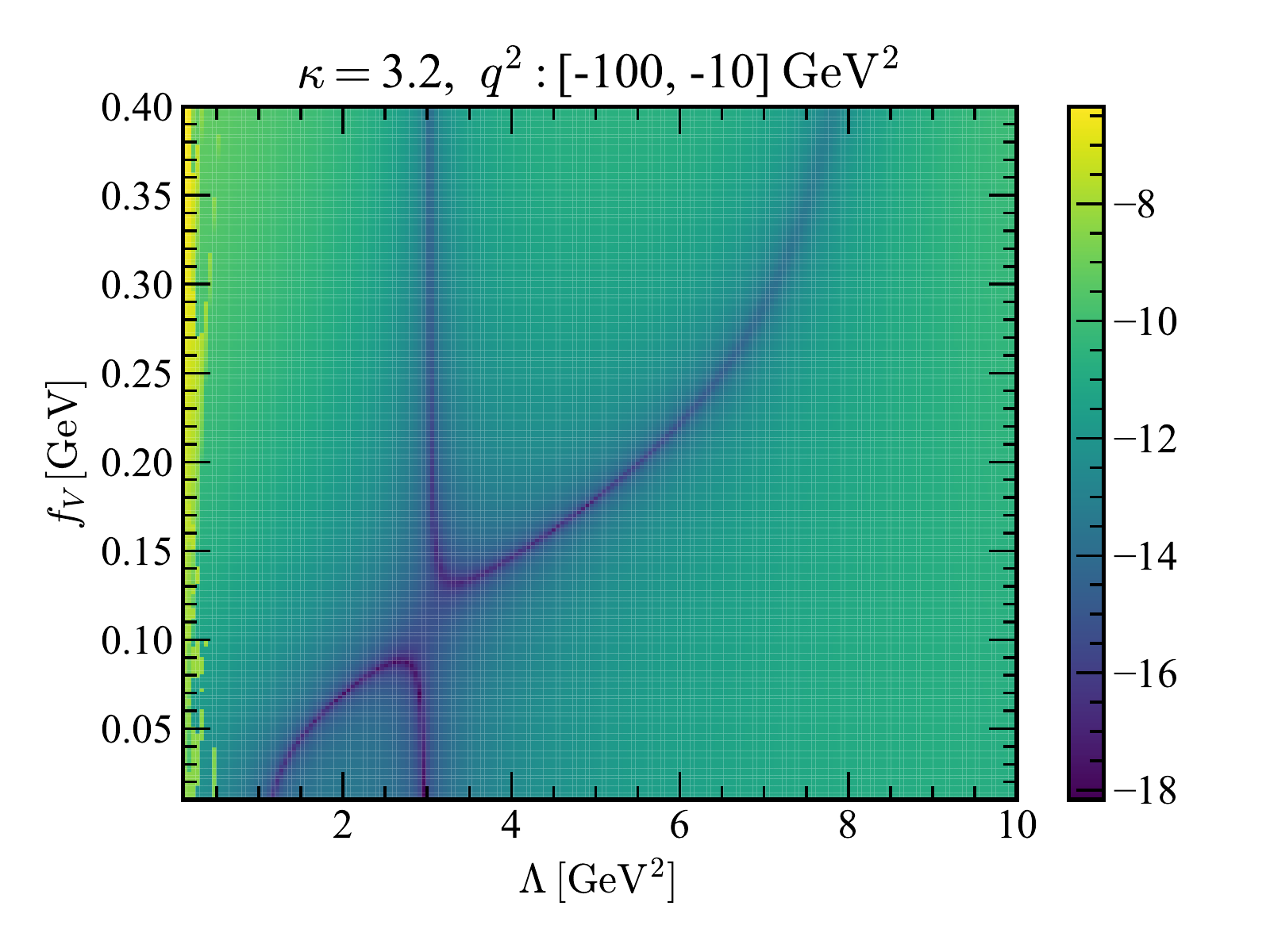}

(a)\hspace{6cm}(b)
\caption{\label{fig7}
Minimum distributions of RSS (a) on the $\Lambda$-$m_V$ plane 
for the double pole parametrization with $\kappa=3.2$ and the input range 
$(-100\;{\rm GeV}^2,-10\;{\rm GeV}^2)$ in $q^2$, and (b) on the $\Lambda$-$f_V$ plane with
$m_V$ being further set to $m_{\rho(1450)}=1.46$ GeV.}
\end{figure}

We observe the RSS minimum distribution on the $\Lambda$-$m_V$ 
plane with the OPE input range $(-100\;{\rm GeV}^2,-10\;{\rm GeV}^2)$ in $q^2$
and $\kappa=3.2$ in Fig.~\ref{fig7}(a). Given the double pole parametrization in 
Eq.~(\ref{imp2}), the minimum distribution indeed becomes less $\Lambda$ dependent 
at large $\Lambda$, compared to the upper minimum distribution in Fig.~\ref{fig3}(a). 
In particular, global minima in the range 
$\Lambda\approx 3$-5 GeV$^2$ imply the preferred values of $m_V$ 
close to the $\rho(1450)$ meson mass $m_{\rho(1450)}\approx 1.46$ GeV. That is, we find
the indication for the existence of the first excited $\rho$ state in our formalism.
We mention that the extraction of higher resonance properties is more sensitive to
the variation of the OPE input, compared to the extraction of the ground state properties:
the $30\%$ variation of the quark condensate $\langle \bar q q\rangle$ causes more than 20\%
difference in the determination of the $\rho(1450)$ meson mass.
A seeming U-shape minimum distribution
attaches to the tilted one without a gap, a layout quite different from Fig.~\ref{fig3}(a). 
Note that there may exist another state $\rho(1570)$ with a similar mass, which has been speculated 
to be due to an Okubo-Zweig-Iizuka-suppressed decay mode of $\rho(1700)$ \cite{Tanabashi:2018oca}. 
It is not clear whether the gapless minimum distributions are related to these two nearby 
states $\rho(1450)$ and $\rho(1570)$. A more precise OPE input may help
clarify this issue.

\begin{figure}
\begin{center}
\includegraphics[width=80mm]{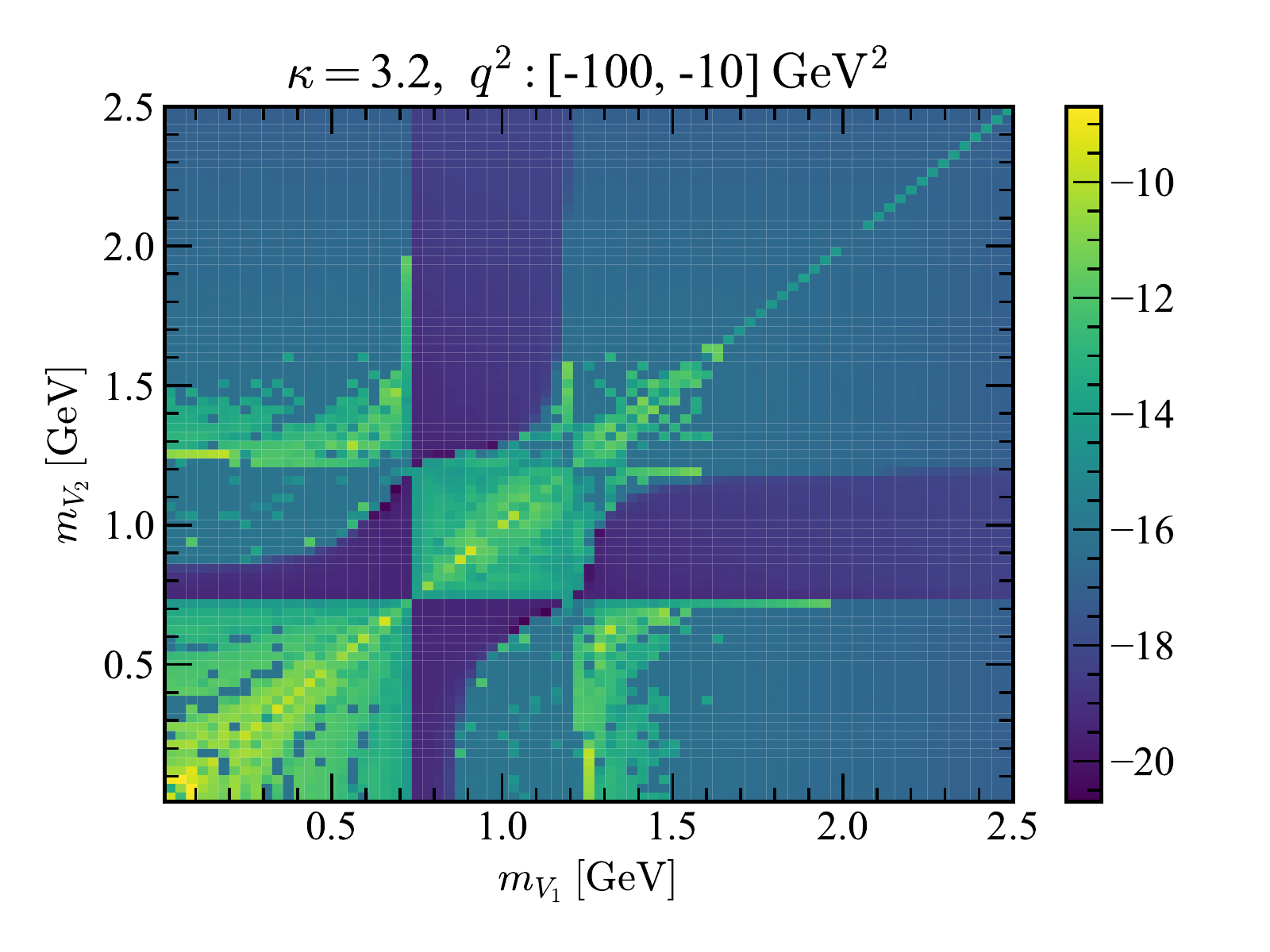}
\caption{RSS minimum distribution on the $m_{V_1}-m_{V_2}$ plane in the double-pole parametrization 
around $\Lambda\approx 2$ GeV$^2$ with $\kappa=3.2$.}\label{m1m2}
\end{center}
\end{figure}

Since extremely low values of RSS have been observed in Fig.~\ref{fig3} for 
the one resonance study, the sum rule is quite stable, and the OPE input
could be described well by one resonance, it is a concern whether we have overfit
the OPE input using the double-pole parametrization in Eq.~(\ref{imp2}). 
To clarify this concern and whether new information
on higher resonances can be extracted, we treat both the masses 
$m_{V_1}$ and $m_{V_2}$, and both the decay constants $f_{V_1}$ and $f_{V_2}$ as free parameters
in a double-pole parametrization, and then scan the $m_{V_1}$-$m_{V_2}$ plane around 
$\Lambda\approx 2$ GeV$^2$ with $\kappa=3.2$ to search for the RSS minimum distribution. 
It is intriguing to see in Fig.~\ref{m1m2} that the region with $m_{V_1}\approx m_{V_2}$, namely,
a single-state solution is not favored compared to the
two-state solution. With more free parameters, the uncertainty
in the fit is larger, as indicated by the wide dark bands, and the numerical fluctuation is more violent. 
It is the reason why we performed the fit and determined the mass one resonance after another
in order to control the precision.
Nevertheless, global minima are still identified in Fig.~\ref{m1m2}, which hint
a state with a lower mass about 0.8 GeV and another state with a higher mass about 1.3 GeV. Therefore, 
nontrivial new information on higher resonances can indeed be extracted from the OPE input 
through the sum rules.

We then choose $m_V$ in Eq.~(\ref{imp2}) as $m_{\rho(1450)}=1.46$ GeV, 
namely, fix the considered excited state to be $\rho(1450)$, and find the minimum 
distribution on the $\Lambda$-$f_V$ plane in Fig.~\ref{fig7}(b). We discard the distribution in the low
$\Lambda$ region, since the separation scale should be higher 
than $m_{\rho(1450)}^2$ in the search for a physical solution of the double pole parametrization. 
The preferred decay constant $f_{\rho(1450)}=0.19$ GeV is read off from 
the global minimum located at $\Lambda= 4.8$ GeV$^2$.
This value of $f_{\rho(1450)}$ leads to the ratio of the two $\tau$ decay widths,
\begin{eqnarray}
\frac{\Gamma(\tau\to \rho(1450)\nu_\tau)}{\Gamma(\tau\to \rho(770)\nu_\tau)}
=\frac{(m_\tau^2-m_{\rho(1450)}^2)f_{\rho(1450)}^2}{(m_\tau^2-m_{\rho(770)}^2)f_{\rho(770)}^2}
\approx 0.3,
\end{eqnarray}
for the masses $m_\tau=1.777$ GeV, $m_{\rho(770)}=0.775$ GeV and $m_{\rho(1450)}=1.465$ GeV,
and the decay constant $f_{\rho(770)}=0.22$ GeV.
The above ratio, being larger than the estimate 0.1 in the extended Nambu-Jona-Lasinio model
\cite{Ahmadov:2015oca}, can be confronted with future data.

We point out that the continuum contribution to the spectral density differs from the
one described by Eq.~(\ref{coe1}), because the first excited state $\rho(1450)$ has been moved out 
of the continuum. The coefficients of the Legendre polynomials corresponding to the global 
minimum in Fig.~\ref{fig7}(b) are
\begin{eqnarray}
b_0=0.0104 , \;\;\;\;b_1= 0.0248,\;\;\;\; b_2=0.0033,\;\;\;\; b_3= -0.0101,\label{coe3}
\end{eqnarray}
so the expansion of the spectral density function up to the $P_3$ term is still enough. The 
behavior of the spectral density function $\rho^h(q^2)$ in $q^2$ is displayed in 
Fig.~\ref{fig4}(a), which differs from the step function in Eq.~(\ref{non}) and from the one
associated with the single pole solution. It is natural that $\rho^h(q^2)$ becomes sizable 
at higher $q^2$, when more resonances are moved out of the continuum. 
Without the duality assumption on the hadron side, there is more freedom to adjust the continuum contribution 
according to considered resonances.

\begin{figure}
\includegraphics[scale=0.40]{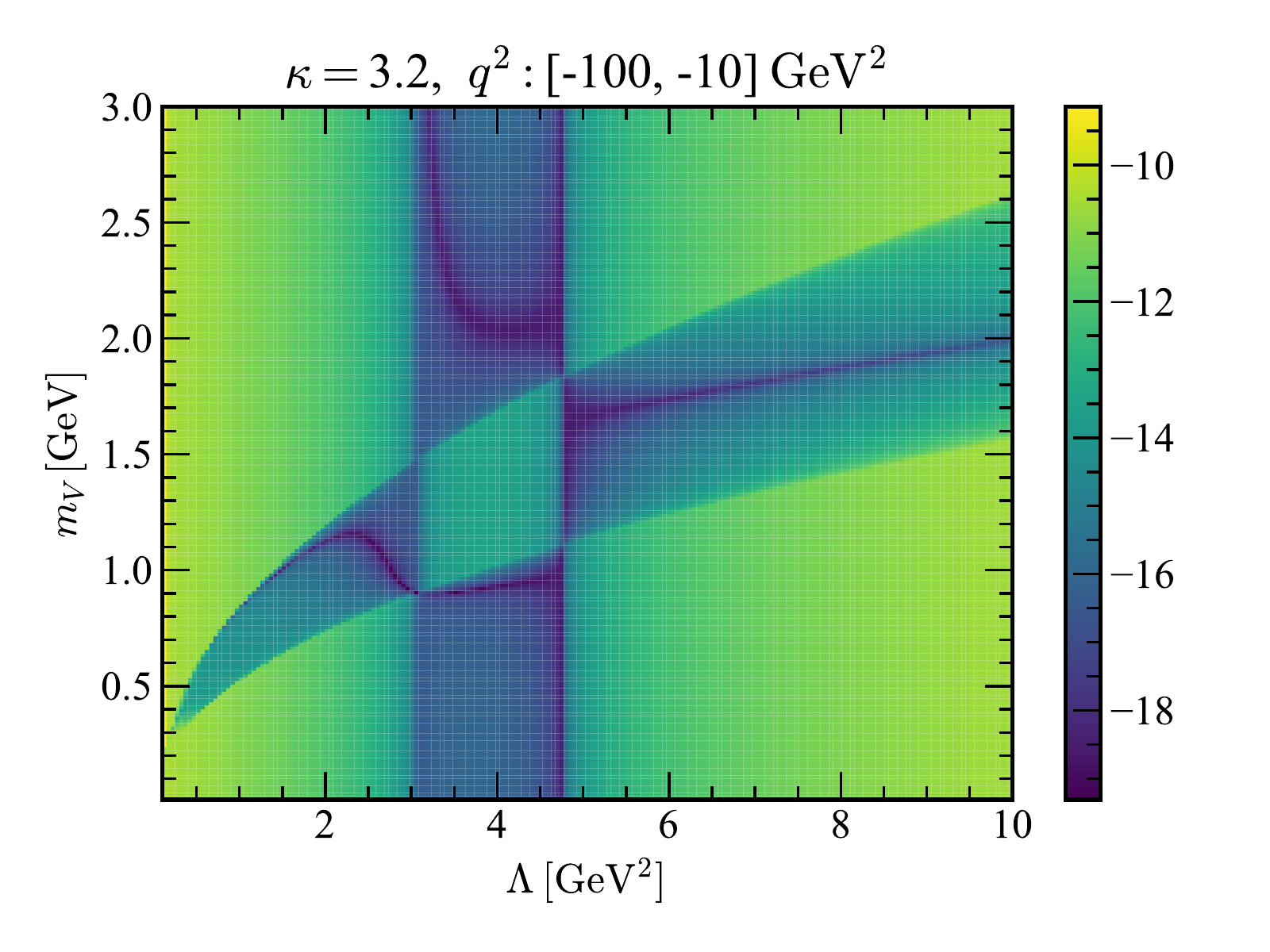}
\includegraphics[scale=0.40]{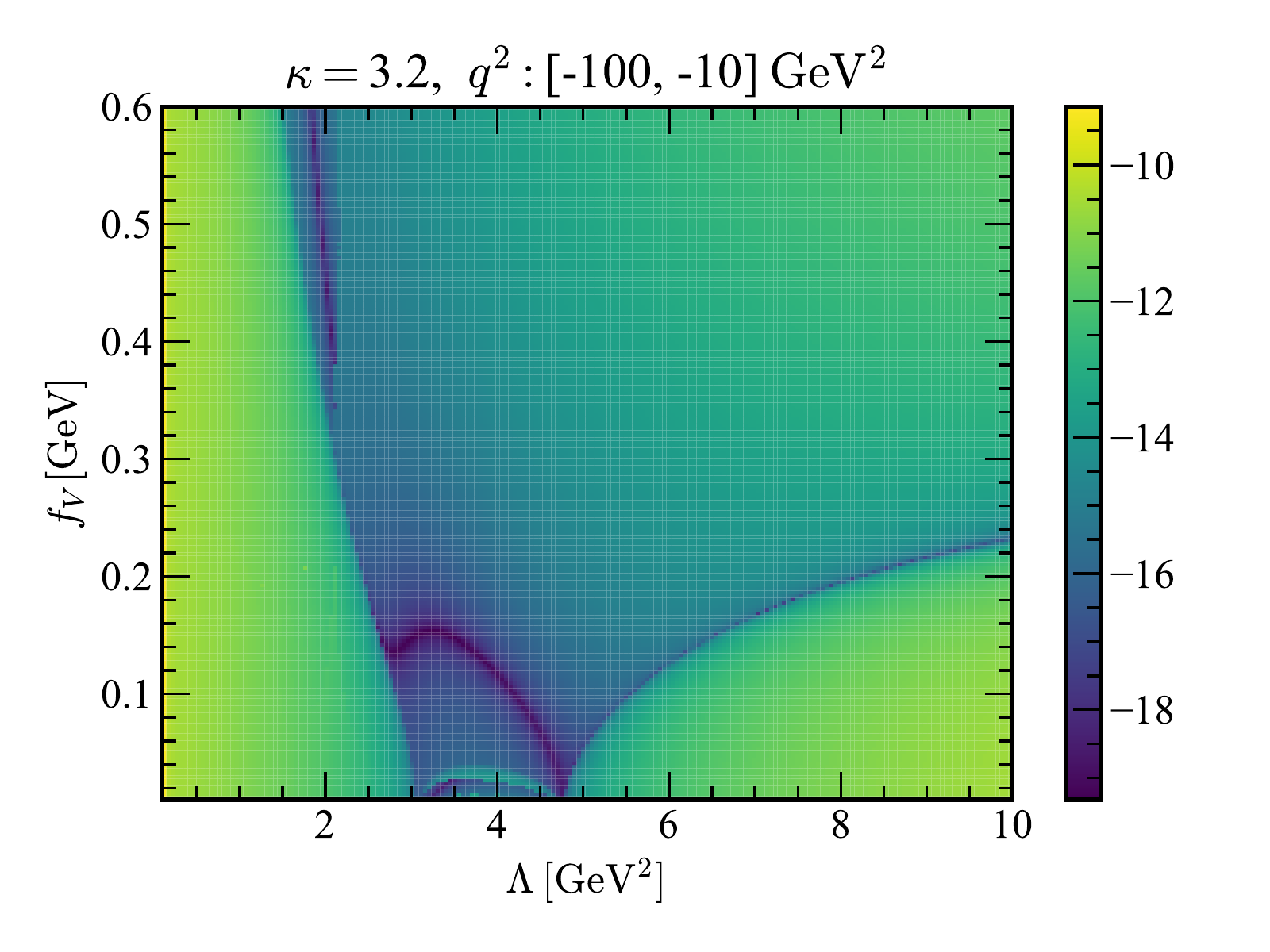}

(a)\hspace{6cm}(b)
\caption{\label{fig9}
Minimum distributions of RSS (a) on the $\Lambda$-$m_V$ plane 
for the triple pole parametrization with $\kappa=3.2$ and the input range 
$(-100\;{\rm GeV}^2,-10\;{\rm GeV}^2)$ in $q^2$, and (b) on the $\Lambda$-$f_V$ plane with
$m_V$ being further set to $m_{\rho(1700)}=1.7$ GeV.}
\end{figure}

The strategy to extract the observables associated with the next excited state is clear 
now in our formalism. The study of higher excited states is expected to be more difficult, 
since they become denser in the mass spectrum. Motivated by the appearance 
of the additional U-shape minimum distribution on the $\Lambda$-$m_V$ plane
in Fig.~\ref{fig9}(a), we repeat the procedure. To examine whether there exist higher 
excited states, we further modify the spectral density into
\begin{eqnarray}
{\rm Im}\Pi(q^2)=\pi f_{\rho(770)}^2\delta(q^2-m_{\rho(770)}^2)+\pi f_{\rho(1450)}^2\delta(q^2-m_{\rho(1450)}^2)
+\pi f_V^2\delta(q^2-m_V^2)+\pi\rho^h(q^2),\label{imp3}
\end{eqnarray}
where the mass and the decay constant of the $\rho(1450)$ meson have been set to
$m_{\rho(1450)}=1.46$ GeV and $f_{\rho(1450)}=0.19$ GeV derived above, respectively. 
The RSS minimum distribution on the $\Lambda$-$m_V$ 
plane with the OPE input range $(-100\;{\rm GeV}^2,-10\;{\rm GeV}^2)$ in $q^2$
and $\kappa=3.2$ is presented in Fig.~\ref{fig9}(a). The U-shape minimum distribution
appears again, but with a gap above the slightly tilted one. It is easy to find the global minima 
located on the tilted minimum distribution around $\Lambda\sim 5$ GeV$^2$, which correspond to 
$m_V\approx 1.7$ GeV, exactly the $\rho(1700)$ meson mass \cite{Tanabashi:2018oca}.
That is, the second excited $\rho$ state also emerges in our formalism.

We then search the minimum distribution on the $\Lambda$-$f_V$ plane with 
$m_V$ being set to the value of $m_{\rho(1700)}=1.7$ GeV, 
namely, with the considered excited state being fixed to $\rho(1700)$. The minimum 
distribution, displayed in Fig.~\ref{fig9}(b), also reveals a nontrivial structure.
We read off  the decay constant $f_{\rho(1700)}=0.14$ GeV from the global minimum located 
at $\Lambda= 5.8$ GeV$^2$.  The associated coefficients in the polynomial expansion are modified into
\begin{eqnarray}
b_0=0.0106 , \;\;\;\;b_1= 0.0244,\;\;\;\; b_2=0.0031,\;\;\;\; b_3= -0.0099,\label{coe4}
\end{eqnarray}
and the resultant behavior of the spectral density function $\rho^h(q^2)$ is displayed in Fig.~\ref{fig4}(a).
As expected, the spectral density function shifts further toward the large $q^2$ region
with one more excited state being moved out of the continuum.

\begin{figure}
\includegraphics[scale=0.40]{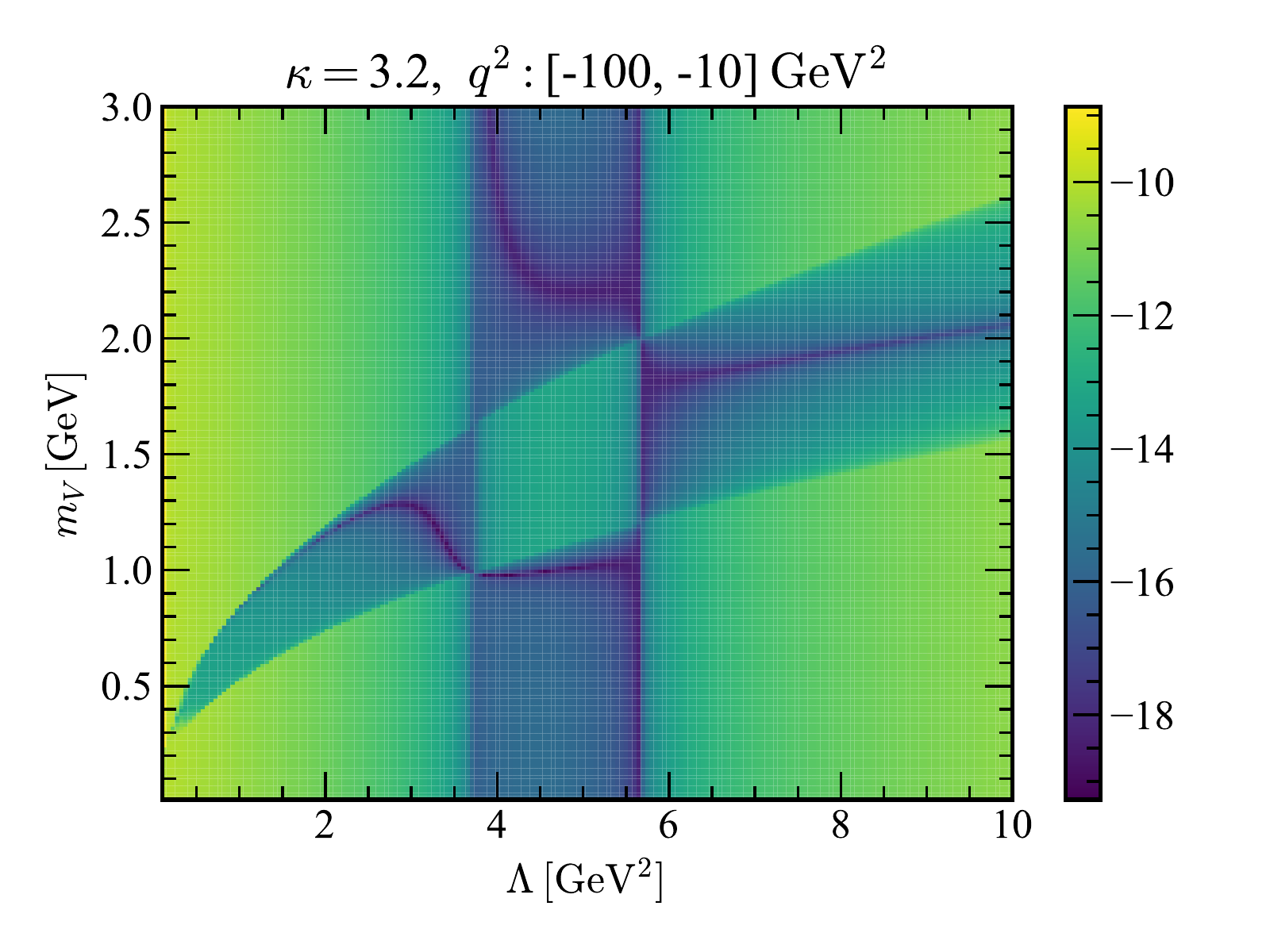}
\includegraphics[scale=0.40]{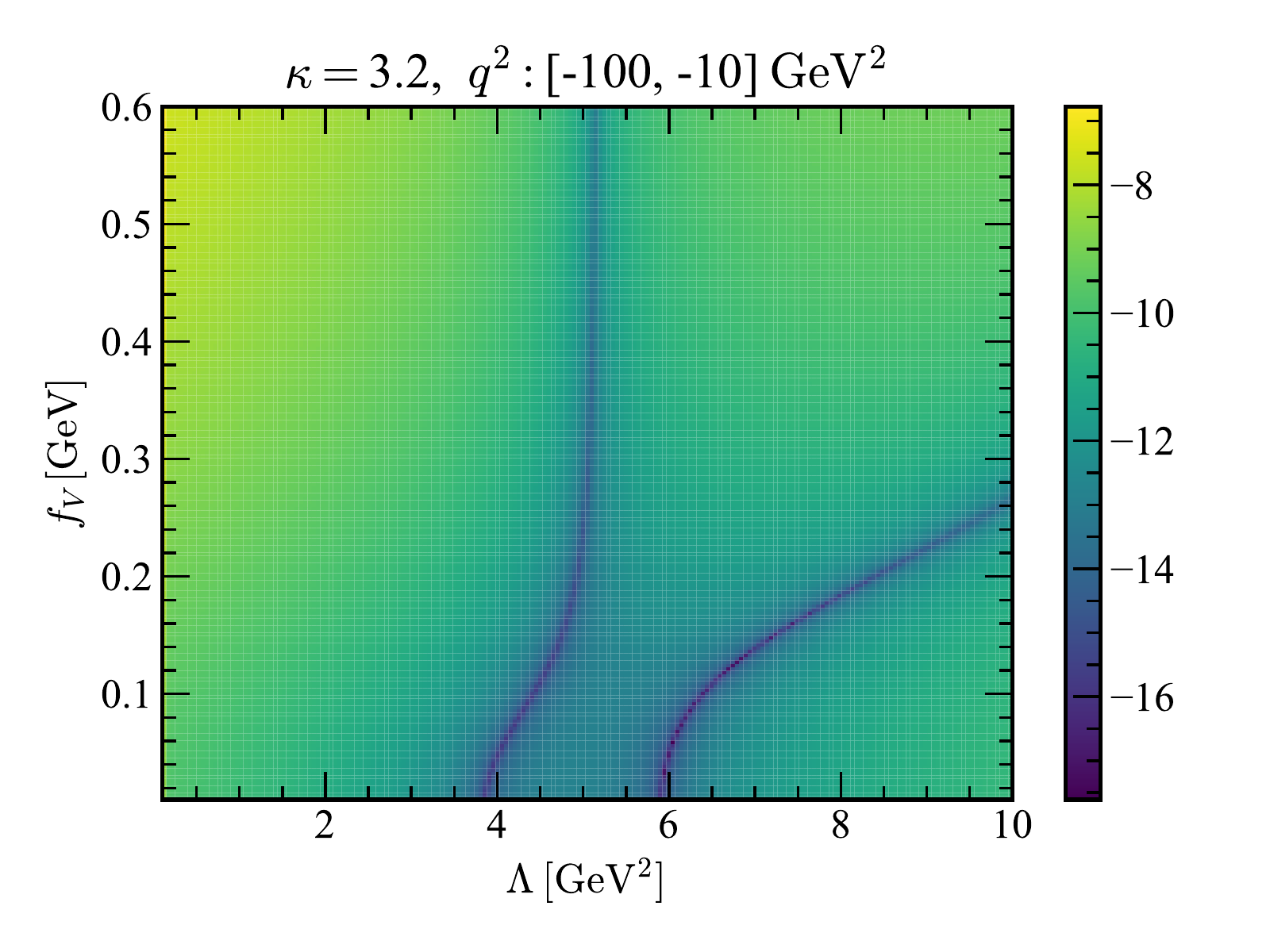}

(a)\hspace{6cm}(b)
\caption{\label{fig10}
Minimum distributions of RSS (a) on the $\Lambda$-$m_V$ plane
for the quadruple pole parametrization with $\kappa=3.2$ and the input range 
$(-100\;{\rm GeV}^2,-10\;{\rm GeV}^2)$ in $q^2$, and (b) on the $\Lambda$-$f_V$ plane with
$m_V$ being further set to $m_{\rho(1900)}=1.9$ GeV.}
\end{figure}

Motivated by the nontrivial U-shape minimum distribution in Fig.~\ref{fig9}(a), we study next 
excited state. Adopting the spectral density
\begin{eqnarray}
{\rm Im}\Pi(q^2)&=&\pi f_\rho^2\delta(q^2-m_\rho^2)+\pi f_{\rho(1450)}^2\delta(q^2-m_{\rho(1450)}^2)
+\pi f_{\rho(1700)}^2\delta(q^2-m_{\rho(1700)}^2)\nonumber\\
& &+\pi f_V^2\delta(q^2-m_V^2)+\pi\rho^h(q^2),\label{imp4}
\end{eqnarray}
we analyze the minimum distribution on the $\Lambda$-$m_V$ plane shown in Fig.~\ref{fig10}(a), 
where the global minima located in the range $\Lambda\sim 6$-7 GeV$^2$ give
$m_V\sim 1.9$ GeV. It is exactly the $\rho(1900)$ meson mass \cite{Tanabashi:2018oca}, implying that
the third excited $\rho$ state still emerges in our formalism.
We then search the minimum distribution on the $\Lambda$-$f_V$ plane with 
$m_V$ being set to $m_{\rho(1900)}=1.9$ GeV, 
namely, with the considered excited state being fixed to $\rho(1900)$. The global minimum 
in Fig.~\ref{fig10}(b) located at $\Lambda= 7.1$ GeV$^2$ gives  
the decay constant $f_{\rho(1900)}=0.14$ GeV, which marks a prediction of
our formalism that has not yet been attempted before. 
The corresponding coefficients in the polynomial expansion are 
\begin{eqnarray}
b_0=0.0118 , \;\;\;\;b_1= 0.0242,\;\;\;\; b_2=0.0028,\;\;\;\; b_3= -0.0095,\label{coe5}
\end{eqnarray}
and the resultant behavior of the spectral density function $\rho^h(q^2)$ is exhibited in 
Fig.~\ref{fig4}(a).

\begin{figure}
\includegraphics[scale=0.40]{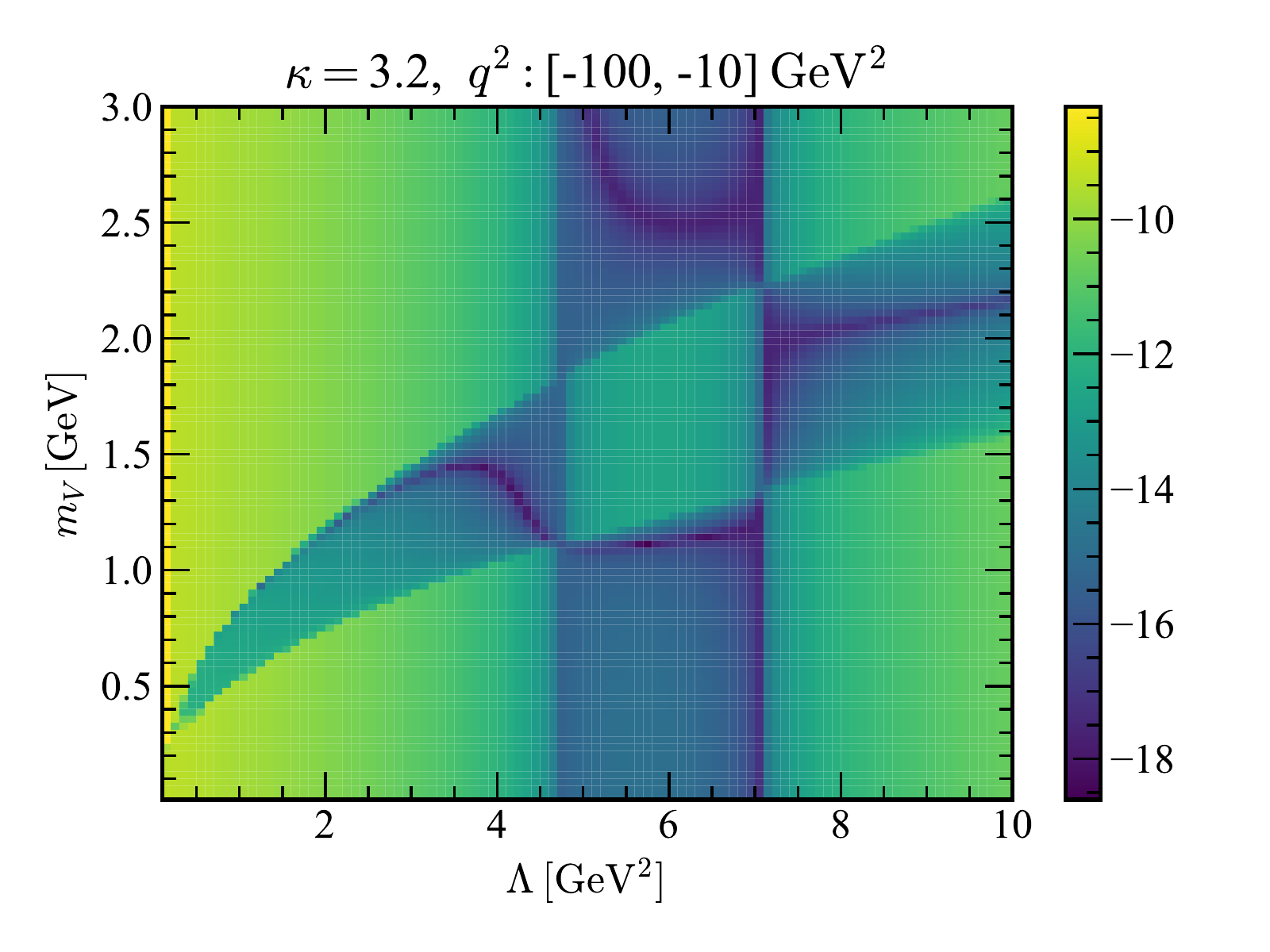}
\caption{\label{fig105}
Minimum distributions of RSS on the $\Lambda$-$m_V$ plane
for the quintuple pole parametrization with $\kappa=3.2$ and the input range 
$(-100\;{\rm GeV}^2,-10\;{\rm GeV}^2)$ in $q^2$.}
\end{figure}

One may wonder whether even higher excited $\rho$ states can be probed in our
formalism, because some nontrivial U-shape minimum distribution still shows up in Fig.~\ref{fig10}(a),
with a small gap above the lower minimum distribution. Above $\rho(1900)$, there are 
$\rho(2150)$, which is a well-established state, and $\rho(2000)$, which is
poorly established and needs confirmation \cite{Tanabashi:2018oca}. We extend the parametrization
in Eq.~(\ref{imp4}) to include one more unknown pole, with the other poles being fixed in the
previous analysis. The scanning on the $\Lambda$-$m_V$ plane reveals the minimum distribution
in Fig.~\ref{fig105}. An obvious global minimum is located at $\Lambda\approx 7$ GeV$^2$ with the 
corresponding mass $m_V\sim 2.0$ GeV, which supports the existence of the $\rho(2000)$
state. Though there is a vague U-shape minimum distribution above the tilted one, we 
will not proceed further the test application, and end the search of the excited $\rho$ 
states here. 

Our results for the decay constants of the $\rho(1450)$ and $\rho(1700)$
excitations are comparable to those derived in the literature, such as
the sum rule analysis with nonlocal consensate corrections \cite{Bakulev:1998pf,Pimikov:2013usa},
the multiple pole QCD sum rules \cite{Pivovarov:1997da,MaiordeSousa:2012vv},
the light cone quark model \cite{Arndt:1999wx}, the lattice QCD \cite{Yamazaki:2001er},
and the rainbow-ladder truncation method \cite{Qin:2011xq}. The theoretical and experimental 
studies on the $\rho(1900)$ and higher states are still rare. Our formalism can be applied to
the extraction of decay widths for excited $\rho$ states in principle, which, however, demands more 
effort. Simply adding one more time-like pion form factor associated with $\rho(1450)$ 
to the $\rho(770)$ contribution,
we observe no minimum distribution on the $\Lambda$-$\Gamma$ plane. This is not a surprise, since
the $\pi^+\pi^-$ production in $e^+e^-$ annihilation is basically saturated by the $\rho(770)$ intermediate
resonance, and $\rho(1450)$ contributes little. At last, we make a remark on the maximum entropy 
method for sum rules. It is unlikely to reveal all the bound states 
simultaneously, especially when the mass spectrum becomes dense and multiple solutions
exist for such an inverse problem. It may be possible to explore excited states
using this method, if one follows our strategy: find the best fit solutions for 
excited states one by one. We will validate this conjecture in a future publication.

\subsection{Sum Rules with Duality Assumption}

\begin{figure}
\includegraphics[scale=0.40]{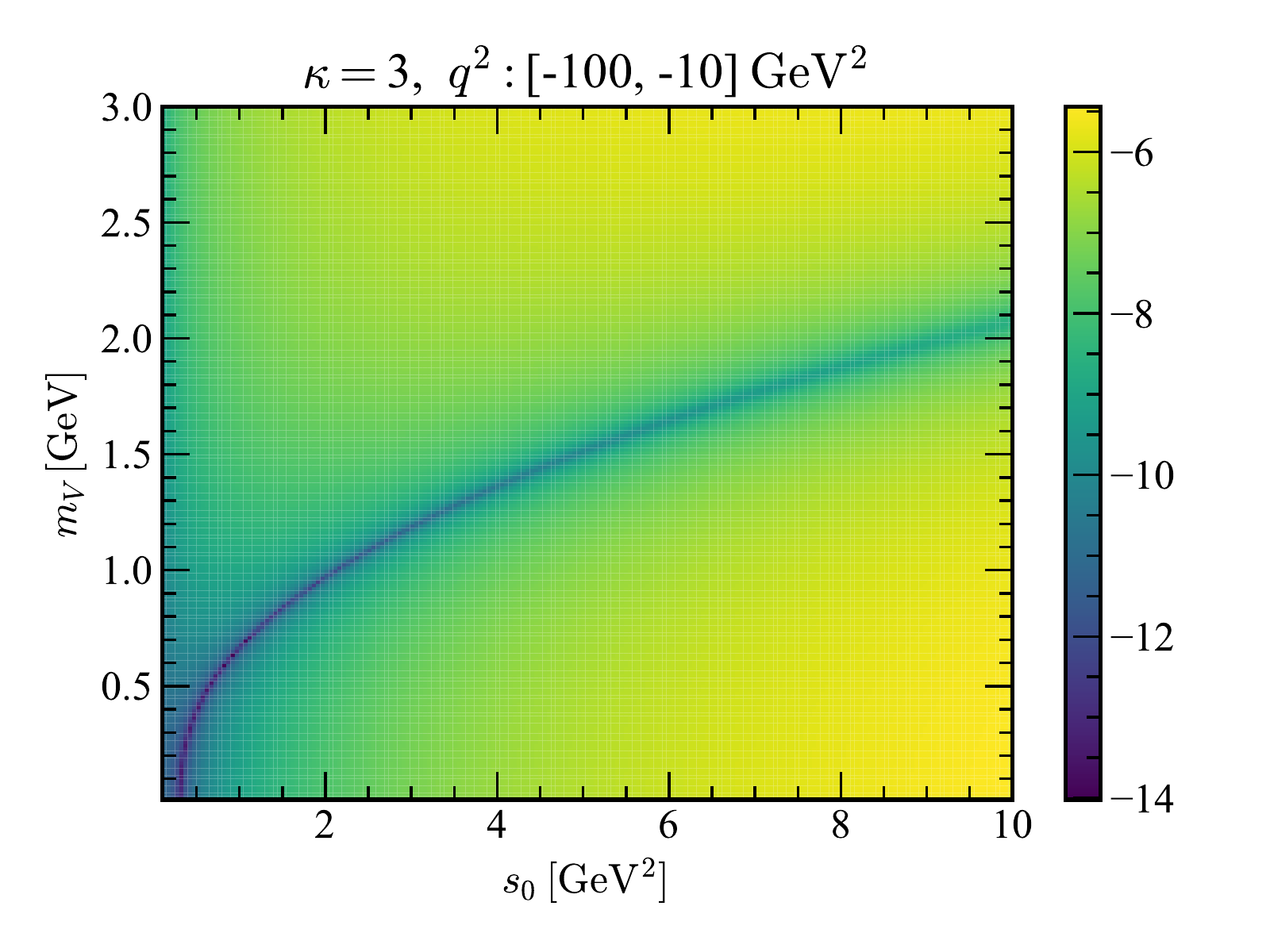}
\includegraphics[scale=0.40]{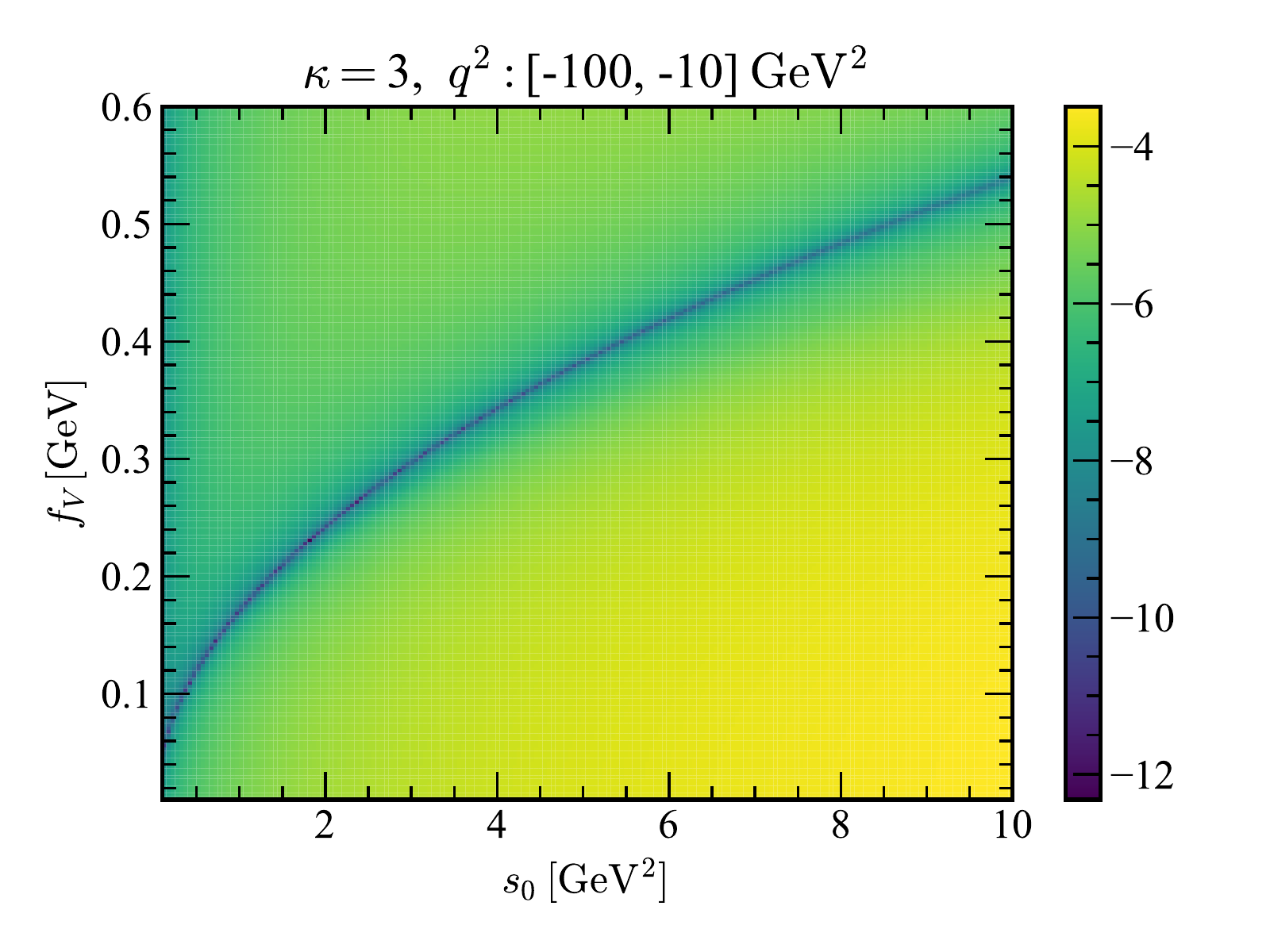}

(a)\hspace{6cm}(b)
\caption{\label{fig11}
Minimum distributions of RSS for the conventional sum rules under the duality
assumption (a) on the $s_0$-$m_V$ plane and  (b) on the $s_0$-$f_V$ plane
with $\kappa=3$ and the input range $(-100\;{\rm GeV}^2,-10\;{\rm GeV}^2)$ in $q^2$.}
\end{figure}

As an alternative viewpoint, the duality assumption in Eq.~(\ref{non}) can be regarded as an 
over-simplified parametrization with a single parameter $s_0$ for the spectral density in 
conventional sum rules. This simple parametrization with a step function satisfies the 
boundary conditions automatically: it vanishes at $s=0$, and the duality assumption guarantees
the continuity condition at $s=s_0$. Compared to our polynomial expansion,
we have two free parameters, $b_0$ and $b_1$. In this sense, our 
parametrization may be simple too, but still more general than the duality assumption. 
We will elaborate that one may not be able to explore properties of excited states reliably 
under the duality assumption. For the convenience of discussion, we present the version 
of Eq.~(\ref{sumb}) before the Borel transformation,
\begin{eqnarray}
\frac{f_V^2}{m_V^2-q^2}=\frac{1}{\pi}\int_{0}^{s_0} ds\frac{{\rm Im}\Pi^{\rm pert}(s)}{s-q^2}+
\frac{1}{12\pi}\frac{\langle\alpha_sG^2\rangle}{(q^2)^2}+
2\frac{\langle m_q \bar q q\rangle}{(q^2)^2} +\frac{224\pi}{81}
\frac{\kappa \alpha_s\langle \bar q q\rangle^2}{(q^2)^3},
\label{sum}
\end{eqnarray}
where the lower bound $s_i$ in the integral on the right hand side has been approximated by zero.
The conventional sum rules in Eqs.~(\ref{sumb}) and (\ref{sum}) can also be 
handled as an inverse problem with the three unknowns $m_V$, $f_V$ and $s_0$. 
This handling is basically the same as the best fit performed in 
\cite{Wang:2016sdt}, and more sophisticated than in \cite{Krasnikov:1981vw,Krasnikov:1982ea},
where the unknowns were solved by requiring the same asymptotic behavior for both sides of the 
sum rules, namely, by equating the coefficients of different powers in $1/q^2$ on both sides. 
We focus only on Eq.~(\ref{sum}), and take
the OPE from the range $(-100\;{\rm GeV}^2,-10\;{\rm GeV}^2)$ in $q^2$ as the input. It has been 
verified that results derived from the sum rule under the Borel transformation in Eq.~(\ref{sumb}) 
are the same. 

The minimum distributions on the $s_0$-$m_V$ and $s_0$-$f_V$ planes for 
$\kappa=3$ are presented in Figs.~\ref{fig11}(a) and \ref{fig11}(b), respectively. The minimum
distributions for $\kappa=2$ and 4 are similar.
It is found that there is only one minimum distribution on the $s_0$-$m_V$, which
increases monotonically with the threshold $s_0$, like the lower minimum distribution in
Fig.~\ref{fig5}(a) attributed only to the perturbative piece of the OPE. 
Note that Fig.~\ref{fig3}(a) contains two minimum 
distributions, where the lower one, corresponding to $m_V\approx 0.78$ GeV,
is stable with respect to the variation of $\Lambda$, and the upper one appears
with a mass gap. The range of the threshold $s_0$ is supposed to be between 
$m_{\rho(770)}^2\approx 0.6$ GeV$^2$ and $m_{\rho(1450)}^2\approx 2.2$ GeV$^2$, within which
neither a global minimum nor a plateau exists around the $\rho(770)$ meson mass. 
Certainly, one can choose an appropriate
$s_0$ value, say, $s_0\approx 1.3$ GeV$^2$ to get $m_V=0.78$ GeV. Then with this $s_0$, 
one can read off $f_V=0.2$ GeV from Fig.~\ref{fig11}(b). In this sense the conventional sum rules are
less predictive, but still useful for estimating the decay constant of 
the ground state, if its mass is fixed. 

\begin{figure}
\includegraphics[scale=0.40]{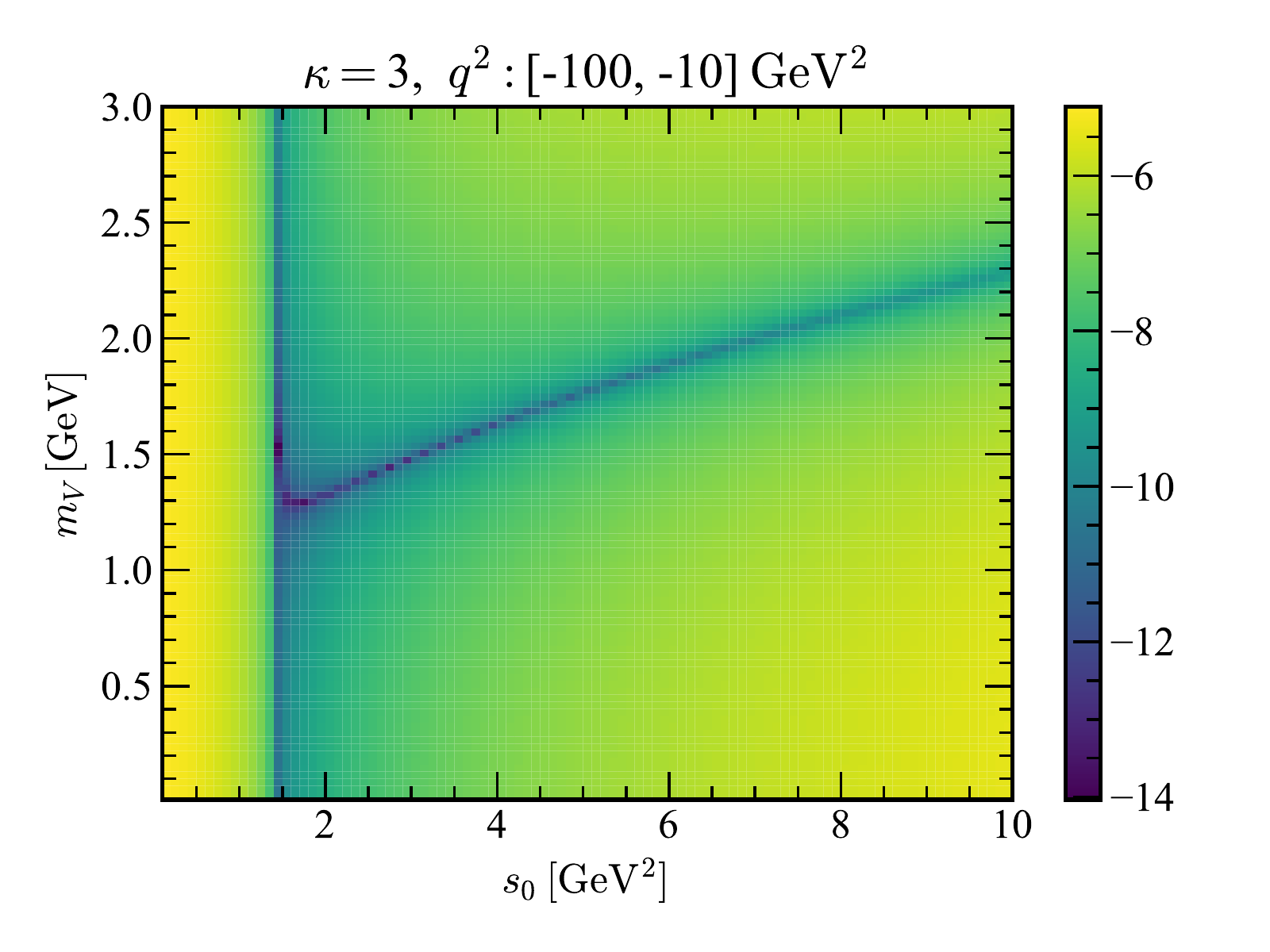}
\includegraphics[scale=0.40]{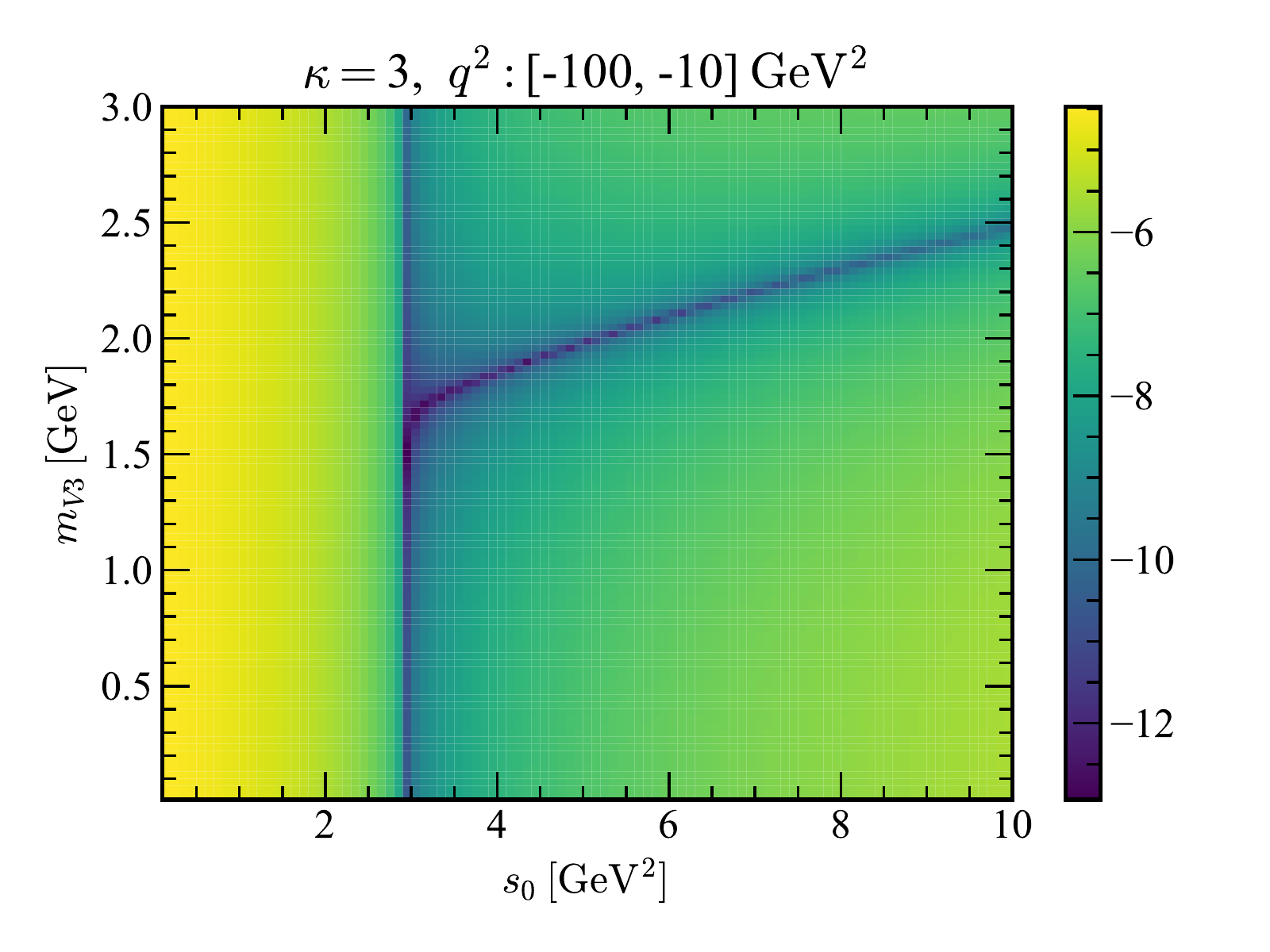}

(a)\hspace{6cm}(b)
\caption{\label{fig12}
Minimum distributions of RSS on the $s_0$-$m_V$ plane for the conventional sum rules 
under the duality assumption (a) for the double pole parametrization, where the first
pole is fixed to be $\rho(770)$, and  (b) for the triple pole parametrization, where the second
pole is further fixed to be $\rho(1450)$
with $\kappa=3$ and the input range $(-100\;{\rm GeV}^2,-10\;{\rm GeV}^2)$ in $q^2$.}
\end{figure}

To check whether the conventional sum rule can probe the $\rho(1450)$ state, we
apply the same strategy, solving the spectral density with a double pole 
parametrization, in which the ground state parameters are fixed to 
$m_{\rho(770)}=0.78$ GeV and $f_{\rho(770)}=0.2$ GeV read off above. Again, the
$s_0$-$m_V$ plane contains only one minimum distribution of RSS as shown in Fig.~\ref{fig12}(a).
In this case a reasonable value of $s_0$ is supposed to be a bit higher than 
$m_{\rho(1450)}^2\approx 2.1$ GeV$^2$ and a bit lower than $m_{\rho(1700)}^2\approx 2.9$ GeV$^2$, 
as the finite widths of these excited states are taken into accounted. No global minimum, ie., 
no reliable solution for a resonance, exists within this range,  
and results of $m_V$ are always below the physical one $m_{\rho(1450)}\approx 1.46$ GeV. The minimum 
at the lower bound of $s_0$ is deeper, but one has to push $s_0$ to the extreme 
$s_0=2.9$ GeV$^2$ in order to barely reach $m_V=1.46$ GeV. The corresponding decay constant 
$f_V=0.21$ GeV, being larger than $f_{\rho(770)}=0.2$ GeV of the ground state, is unlikely
\cite{MaiordeSousa:2012vv}. This is not a surprise, because $s_0=2.9$ GeV$^2$ is not a 
reasonable choice. If one insists on continuing to probe the next excited $\rho$ state, the triple
pole parametrization can be adopted, with the parameters of the second pole being
further fixed to $m_{\rho(1450)}=1.46$ GeV and $f_{\rho(1450)}=0.21$ GeV. The resultant 
minimum distribution on the $s_0$-$m_V$ plane is exhibited in Fig.~\ref{fig12}(b).
It is seen that there is no global minimum, and all values of $m_V$ in the designated range of 
$s_0> (m_{\rho(1700)}+0.1)^2\approx 3.2$ GeV$^2$ \cite{MaiordeSousa:2012vv}
with the finite width of $\rho(1700)$ being considered, are all greater than 1.7 GeV.
Namely, no reliable and sensible solution is identified for the $\rho(1700)$ state.

The above investigation reveals clearly the limitation of conventional sum rules based on
the duality assumption. Though nontrivial minimum distributions of RSS still appear on
the $s_0$-$m_V$ plane, which are allowed by an ill-posed inverse problem, the duality assumption 
imposes too strong restriction on the shape of the continuum. It has to be a step
function with the height being equal to that of the perturbative spectral density,
and the threshold $s_0$ exists in a narrow interval. It means that the continuum
has been roughly fixed, especially as excited states are probed, which are
denser in the mass spectrum. Under this stringent restriction, solutions for resonances
may not exist due to the absence of global minima, and can hardly be correct, even as
$s_0$ is stretched unreasonably. In summary, conventional sum rules may work 
for ground state studies, in which the interval of $s_0$, namely, the flexibility
of varying the continuum is bigger, but should become unreliable for
excited states.

\section{CONCLUSION}

In this paper we have improved QCD sum rules for nonperturbative studies without  
assuming the quark-hadron duality on the hadron side. 
The spectral density at low energy, 
including both resonance and continuum contributions, is solved with the OPE input 
on the quark side by treating sum rules as an inverse problem.
We have elaborated the postulation that the Borel transformation is not crucial for this
new formalism, because the continuum contribution needs not to be suppressed, but is solved via
the inverse problem, and the convergence of the OPE is achieved by adopting
the input in the deep Euclidean region. Once the unknown spectral density is
solved directly, the stability criterion for conventional sum rules is not necessary either. 
The implementation of the above formalism has been demonstrated by identifying the
series of $\rho$ states and by determining their corresponding decay constants from the two-current
correlator. The strategy is to include resonances one by one into the spectral density with
different associated continuum contributions, and to repeat
solving the sum rules by minimizing the difference between the hadron and quark sides.
One should make sure in the above procedure that the scale $\Lambda$, separating the
perturbative and nonperturbative regimes, should be above the highest resonance parametrized  
into the unknown spectra density for consistency. In this way we have predicted 
the decay constants $f_{\rho(770)} (f_{\rho(1450)},f_{\rho(1700)},f_{\rho(1900)})\approx$ 
0.22 (0.19, 0.14, 0.14) GeV
for the masses $m_{\rho(770)} (m_{\rho(1450)},m_{\rho(1700)},m_{\rho(1900)})\approx$ 0.78 (1.46, 1.7, 1.9) GeV
of the $\rho$ resonances. The decay width $\Gamma_{\rho(770)}\approx 0.17$ 
GeV of the $\rho(770)$ meson has been also obtained. We mentioned that the existence of the $\rho(1570)$ state 
could not be excluded, which has been speculated to be due to an Okubo-Zweig-Iizuka-suppressed decay mode 
of $\rho(1700)$, and that the existence of the $\rho(2000)$ state is supported.

The major sources of theoretical uncertainties arise from the OPE for the OPE inputs,
which is truncated at finite orders in $\alpha_s$ and at finite powers of $1/q^2$. 
We have observed that the variation of the factorization violation parameter $\kappa$
from 3 to 4 causes about 10\% variation to the results presented in this work.
More precise inputs, such as reliable $\kappa$ values and condensates, help
determine nonperturbative observables. Here we have fixed $\kappa$ to be 3.2 to
produce the $\rho(770)$ meson mass. The dimension-eight
condensates are still quite uncertain \cite{Boito:2012nt, Blok:1997yd,GPP,Dominguez:2014fua}, and deserve
more investigation. Second, the expansion of the spectral density
function $\rho^h$ in a series of Legendre polynomials is truncated at the fourth
term. Though the convergence of this expansion has been scrutinized, the precision of our 
predictions can be improved by including higher order polynomials. When this is done, the
continuity of the slope of the spectral density function at the separation scale can also 
be imposed, and its impact is worth investigation. It is claimed that
both the above sources of theoretical uncertainties can be reduced straightforwardly and systematically.
One may still question whether the separation scales $\Lambda$ about few GeV$^2$ in our study are large enough
for justifying the replacement of the continuum contribution ${\rm Im}\Pi(s)$ by the 
perturbative one ${\rm Im}\Pi^{\rm pert}(s)$, which is also based on the quark-hadron
duality. Certainly, it is not the concerned duality assumption in conventional QCD
sum rules around the threshold $s_0\approx 1$ GeV$^2$, and the duality violation above
$\Lambda\approx$ few GeV$^2$ is expected to be minor.

This new formalism is more predictive with less ambiguity and
more control of theoretical uncertainty, compared to conventional sum rules, because
the unknown observables were extracted from the best fit of the two sides of a sum rule. In particular,
it can be extended to analyses of excited states, which are difficult to achieve in conventional 
sum rules. We have observed that the condensate corrections up to dimension-six seem to be
sufficient for generating most known $\rho$ resonances.
Since whether a bound state exists can be explored in our formalism, it would be 
of interest to apply it to the various exotic channels, such as those containing more than three quarks. 
It is worthwhile to generalize it to pursue nonperturbative 
properties of vector mesons in nuclear medium \cite{Hatsuda:1991ez, Leupold:1997dg} at finite density or 
temperature to directly observe the change in the spectral density function in hot or dense 
environments. Our formalism cannot 
only be applied to low energy light flavor processes, but also to heavy flavor physics 
\cite{MN,Belyaev:1993wp,Ball:1992ih,Huang:1996pz,Dai:1997df}. 
It is also possible to extend it to studies of nonlocal condensate effects on 
excited states \cite{Bakulev:1998pf,Pimikov:2013usa}, and on other 
more complicated QCD processes \cite{MR,Grozin:1994hd,BPS,Hsieh:2009qj,Stefanis:2015qha}.
There is no doubt that there are broad applications of our nonperturbative formalism.

\vskip 1.0cm
{\bf Acknowledgement}

This work was supported in part by
the Ministry of Science and Technology of R.O.C. under Grant No.
MOST-107-2119-M-001-035-MY3.


\end{document}